\documentclass[twocolumn,showpacs,preprintnumbers,amsmath,amssymb,superscriptaddress,nofootinbib,english]{revtex4-1}
\usepackage{aas_macros}

\usepackage{times,amsmath,amsfonts,amssymb,epstopdf,array}
\usepackage{graphicx}
\usepackage{dcolumn}
\usepackage{bm}
\usepackage{epsfig}
\usepackage{graphicx}
\usepackage{hyperref}
\usepackage{url}
\usepackage[normalem]{ulem}
\usepackage[T1]{fontenc}
\usepackage{parskip}

\newcommand{\bq}{\begin{eqnarray}}
\newcommand{\eq}{\end{eqnarray}}
\newcommand{\bes}{\begin{subequations}}
\newcommand{\ees}{\end{subequations}}

\newcommand{\tcr}{\textcolor{black}}

\usepackage[usenames,dvipsnames]{color}
\usepackage[normalem]{ulem}
\usepackage{enumerate}

\def\hmpc{h^{-1}\,{\rm Mpc}}

\begin{document}

\title{Testing the quasi-static approximation in $f(R)$ gravity simulations}

\author{Sownak Bose}
\email[Email: ]{sownak.bose@durham.ac.uk}
\affiliation{Institute for Computational Cosmology, Department of Physics, Durham University, Durham DH1 3LE, U.K.}

\author{Wojciech A. Hellwing}
\affiliation{Institute for Computational Cosmology, Department of Physics, Durham University, Durham DH1 3LE, U.K.}
\affiliation{Interdisciplinary Centre for Mathematical and Computational Modelling, University of Warsaw, ul. Pawi\'nskiego 5a, Warsaw, Poland}

\author{Baojiu Li}
\affiliation{Institute for Computational Cosmology, Department of Physics, Durham University, Durham DH1 3LE, U.K.}

\begin{abstract}
Numerical simulations in modified gravity have commonly been performed under the quasi-static approximation -- that is, 
by neglecting the effect of time derivatives in the equation of motion of the scalar field that governs the fifth force 
in a given modified gravity theory. To test the validity of this approximation, we analyse the case 
of $f(R)$ gravity beyond this quasi-static limit, by considering effects, if any, these terms have in the matter and 
velocity divergence cosmic fields. To this end, we use the adaptive mesh refinement 
code {\sc ecosmog} to study three variants ($|f_{R}|= 10^{-4}[$F4$], 10^{-5}[$F5$]$ and $10^{-6}[$F6$]$) of the Hu-Sawicki $f(R)$ gravity model, each of which refers to a different magnitude 
for the scalar field that generates the fifth force. We find that for F4 and F5, which show stronger deviations from 
standard gravity, a low-resolution simulation is enough to conclude that time derivatives make a negligible contribution 
to the matter distribution. The F6 model shows a larger deviation from the quasi-static approximation, but one 
that diminishes when re-simulated at higher-resolution. We therefore come to the conclusion that 
the quasi-static approximation is valid for the most practical applications in $f(R)$ cosmologies. 
\end{abstract}

\maketitle

\section{Introduction}
\label{Intro}
In recent years, theories of modified gravity have become a subject of great interest in alternative approaches modelling the observed acceleration of the Universe \cite{1998AJ....116.1009R,  1999ApJ...517..565P}. Einstein's theory of General Relativity (GR) has been the \tcr{underlying} gravity theory in the standard 
cosmological model of $\Lambda$CDM, the dark energy ($\Lambda$) and (cold) dark matter (CDM) components of which remain unresolved challenges to cosmologists. Modified gravity seeks to answer this question by modifying the theory of gravity itself, most routinely with the addition of scalar, vector or tensorial modifications to the Einstein-Hilbert action that governs GR (see \cite{2012PhR...513....1C} for a comprehensive review).  Of course, one cannot deny the undoubted success of GR in passing local \tcr{and} Solar System tests of gravity, and so it is necessary for any reasonable modified gravity theory to also do the same. \tcr{One} process by which a modified theory reduces to GR on small scales is 
known as {\it screening} \cite{2010arXiv1011.5909K}, of which there are three main types: chameleon \cite{2007PhRvD..75f3501M}, Vainshtein \cite{2000PhLB..485..208D} and \tcr{dilaton/symmetron} screening \cite{2010PhRvL.104w1301H,2010PhRvD..82f3519B}, with different theories equipped with different screening mechanisms.

One of the most popular models of modified gravity is 
$f(R)$ gravity \cite{2005PhRvD..71f3513C}. This theory is built around the addition of a scalar function of the Ricci curvature scalar to the Einstein-Hilbert action. The scalar field \tcr{has a potential, which acts as an effective cosmological constant that accelerates the expansion of the Universe, and also generates a `fifth force' between matter particles.}
\tcr{While the fifth force enhances the standard Newtonian gravity} in low-density regions, 
in high-density regions, GR is 
recovered by means of the chameleon screening. This mechanism is \tcr{a consequence}  of the high degree of non-linearity in the equations of motion that govern this theory. Its presence makes standard perturbative approaches less useful, and calls for the need to perform $N$-body simulations at high-resolution \tcr{to fully understand the cosmological behaviour of this model}.

Numerical simulations 
\tcr{for $f(R)$ gravity (and for most other modified gravity theories)} have traditionally been performed in what is known as the ``quasi-static limit'', \tcr{in which the time derivatives} of the scalar field that generates the fifth force are considered small compared to its spatial derivatives, and can therefore be safely 
\tcr{neglected} \cite{2012JCAP...01..051L, 2014MNRAS.439.2978C, 2013MNRAS.435.2806H}. An advantage of this approximation is that it \tcr{considerably} simplifies the 
challenge of numerically solving the non-linear equations. In GR \tcr{simulations}, this approximation has been tested as being valid, but while it is consistently made in the case of $f(R)$ simulations, 
\tcr{its validity has not yet} been tested rigorously, \tcr{especially} in the non-linear regime \tcr{(we note that recently} efforts to include non-static effects have been made in the case of symmetron screening \cite{2014PhRvD..89h4023L}\tcr{)}.


\tcr{The aim} of our investigation here \tcr{is to quantitatively estimate the effects of excluding the time derivatives in $N$-body simulations for $f(R)$ gravity. For this purpose, we have derived field equations in which time derivatives of the scalar field are consistently included, and implemented these equations in a modified version of the {\sc ecosmog} code \cite{2012JCAP...01..051L}. By running simulations at different resolutions, we then study how the clustering of matter is affected by the non-static effects. We find that in low-resolution simulations, the time derivatives do have an impact on the observables we study, but this diminishes when we re-simulate at higher resolution or shorter time steps. As a result, at least for the $f(R)$ models we have studied, the quasi-static approximation seems to be valid for the observables we are interested in.}


This paper is organised as follows: in Section~\ref{fRintro}, we introduce the \tcr{Hu-Sawicki \cite{2007PhRvD..76f4004H} $f(R)$ model}, and how chameleon screening is able to recover GR. Sections~\ref{freqs} and~\ref{ECO} describe how we modify the ordinary evolution equations to account for time derivatives in the non-linear regime, and how these equations are then discretised for the purpose of solving them on a mesh. 
In Section~\ref{Results}, we present the results of our $N$-body simulations at different resolutions, while in Section~\ref{NumCons}, we discuss some numerical aspects that must be taken into account when interpreting the results of our work. Finally, in Section~\ref{Summary}, we summarise our findings and their implications. 

\tcr{Throughout this paper}, Greek indices run over $0,1,2,3$ (the four space-time components) whereas Latin indices run over $1,2,3$ (the three spatial components). 

\section{An Introduction to $f(R)$ Gravity}
\label{fRintro}

In this section, we will briefly discuss the main features of $f(R)$ gravity, first in general, and then with the more specific example of the Hu-Sawicki \cite{2007PhRvD..76f4004H} model, which is the one we will analyse 
\tcr{in the rest} of this paper. We expect that our findings in this work are \tcr{are at least qualitatively} applicable to other classes of $f(R)$ models as well.

\subsection{$f(R)$ Gravity: an Overview}
\label{fRoverview}

As with most modified gravity theories, the starting point is the Einstein-Hilbert action. The modification we make is to replace the cosmological constant $\Lambda$ with a function of the Ricci scalar, $R$, as:

\bq \label{genfR}
S = \int \mathrm{d}^4 x \sqrt{-g} \left[ \frac{1}{2} M_{\mathrm{Pl}}^2 \left[R + f(R)\right] + \mathcal{L}_m  \right]\;,
\eq 
where $g$ is the determinant of the metric tensor $g_{\mu \nu}$, $M_{\mathrm{Pl}} = 1/\sqrt{8 \pi G}$ is the reduced Planck mass, $G$ is the Newtonian gravitational constant, and $\mathcal{L}_m$ the total matter (baryonic + dark matter) Lagrangian density. \tcr{We} assume that neutrinos are massless, and that at late times the contribution \tcr{from} photons and neutrinos is negligible. The distinction between 
\tcr{different $f(R)$ models is in the specific choice for the function $f(R)$ itself}.

By varying the action in Eq.~\ref{genfR} with respect to the metric $g_{\mu \nu}$, we obtain the modified Einstein field equations:
\bq \label{EFE}
&&G_{\mu \nu} + f_R R_{\mu \nu} - \left[ \frac{1}{2} f (R) - \Box f_R \right] g_{\mu \nu} - \nabla_\mu \nabla_\nu f_R \nonumber \\&&= 8 \pi G T_{\mu \nu}^{\mathrm{m}}\;,
\eq

where $G_{\mu \nu} = R_{\mu \nu} - \frac{1}{2}g_{\mu \nu} R$ is the Einstein tensor, $\nabla_{\mu}$ is the covariant derivative \tcr{compatible with the metric $g_{\mu\nu}$}, $\Box \equiv \nabla^\mu \nabla_\mu$, $T_{\mu \nu}^{\mathrm{m}}$ is the energy-momentum tensor for matter, and $f_R \equiv \frac{\mathrm{d}f(R)}{\mathrm{d}R}$ is the extra scalar degree of freedom \tcr{of this model}, known as the {\it scalaron}. One can straightforwardly obtain the equation of motion for \tcr{the scalar field} by taking the trace of Eq.~\ref{EFE}:

\bq \label{FREOM}
\Box f_R = \frac{1}{3} \left( R - f_R R + 2 f(R) + 8 \pi G \rho_m \right) \;,
\eq
in which $\rho_m$ is the matter density in the Universe. Since we are interested in the cosmological properties of these models, we need to derive the perturbation equations. In order to do this, we will work in the Newtonian gauge:
\bq \label{Ngauge}
\mathrm{d}s^2 = (1 + 2\Psi)\mathrm{d}t^2 - a^2(t)(1 - 2\Phi)\mathrm{d}\vec{x}^2\;,
\eq
where $\Psi$ and $\Phi$ are the gravitational potentials, with $\Psi \neq \Phi$ for the time being (non no-slip condition), $t$ is the \tcr{physical} time, $\vec{x}$ is the comoving coordinate, and $a$ is cosmic scale factor, with $a = 1$ today. 
\tcr{The perturbation is around} the standard Friedmann-Robertson-Walker (FRW) metric, \tcr{which} describes the background evolution of the Universe \tcr{(or of $a(t)$)}. Given this, we can then write down the scalaron equation of motion:

\bq \label{scEOM}
\frac{1}{a^2} \vec{\nabla}^2 f_R \approx -\frac{1}{3} \left[R - \bar{R} + 8 \pi G \left(\rho_m - \bar{\rho}_m \right) \right]\;,
\eq

and the modified Poisson equation:

\bq \label{MPE}
\frac{1}{a^2} \vec{\nabla}^2 \Phi \approx \frac{16 \pi G}{3}\left(\rho_m - \bar{\rho}_m\right) + \frac{1}{6} \left( R - \bar{R}\right)\;,
\eq
where quantities with an overbar signify those defined in the background cosmology, and $\vec{\nabla}$ denotes the three-dimensional spatial derivative \tcr{with respect to $\vec{x}$}. 

\tcr{When deriving Eqs.~\ref{scEOM} and \ref{MPE}, we have assumed that $\left|f(R)\right| \ll \left|R\right|$ and $\left|f_R\right| \ll 1$, which is true for the models we study below. Eqs.~\ref{scEOM} and \ref{MPE} are solved by the standard {\sc ecosmog} code, in which the quasi-static approximation has been used and time derivatives of the scalaron field $f_R$ are neglected. We will show below how to extend these equations consistently to restore those time derivatives.}

\subsection{The Chameleon Screening Mechanism}
\label{chameleon}


\tcr{While} modifying the theory of gravity to \tcr{explain} the \tcr{accelerated expansion} of the Universe on a cosmological level, one must bear in mind the tremendous 
\tcr{success of GR in Solar System tests}. $f(R)$ gravity \tcr{incurs} a fifth force that enhances gravity on large scales, which needs to be \tcr{suppressed} locally to pass those experimental tests. For this reason, \tcr{viable} $f(R)$ models are equipped with a mechanism \tcr{to ensure} that: (1)~gravity is modified (enhanced) \tcr{on cosmological scales}, and (2) GR is recovered in Solar or similar systems. This is known as the {\it chameleon mechanism}. 

To see how this is manifest in $f(R)$ gravity, we can construct an effective potential for the scalaron field as:

\bq \label{effPot}
\frac{\mathrm{d}V_{\mathrm{eff}} \left[f_R ; \rho_m \right) }{\mathrm{d} f_R} = \tcr{-}\frac{1}{3} \left[R - f_R R + 2 f(R) + 8 \pi G \rho_m \right].
\eq

In regions of high matter density ($\rho_m \gg \bar{\rho}_m$), \tcr{$\left|f_R\right| \ll \left|\bar{f}_R\right| \ll 1$}, and so the GR solution $R = -8 \pi G \rho_m$ minimises Eq.~\ref{effPot}, giving rise to an effective mass for the scalaron field:

\bq \label{scaMass}
m_{\mathrm{eff}}^2 \equiv \frac{\mathrm{d}^2V_{\mathrm{eff}}}{\mathrm{d} f_R ^2} \approx \tcr{-}\frac{1}{3} \frac{\mathrm{d} R}{\mathrm{d} f_R} > 0\;.
\eq

This fifth force is Yukawa-type, and decays as $\exp\left( - m_{\mathrm{eff}} r \right)$, where $r$ is the separation between two test masses. 
\tcr{According to} Eq.~\ref{effPot}, 
$m_{\mathrm{eff}}$ depends explicitly on $\rho_m$, \tcr{and} we can see from Eq.~\ref{scaMass} that in regions of high matter density (or equally, where the Newtonian potential is deep), the fifth force is more strongly suppressed \tcr{as $m_{\rm eff}$ is larger there (which is because $|R| \approx 8\pi G\rho_m$ is large and $|f_R|$ small in high-density regions)}. The deviations from GR become practically undetectable, and hence the GR limit is 
recovered in those regimes.

\subsection{The Hu-Sawicki Model}
\label{HuSawicki}

Thus far, the discussion has been quite general, without specifying the functional form for $f(R)$. Note that the choice for the form of $f(R)$ completely specifies the model. The Hu-Sawicki model is one such example, which takes the following form:

\bq \label{HuS}
f(R) = - M^2 \frac{c_1 \left(-R/M^2\right)^n}{c_2 \left(R / M^2\right)^n + 1}\;,
\eq

where $M$ is a characteristic mass scale, defined by $M^2 = 8 \pi G \bar{\rho}_{\mathrm{m}0} / 3 = H_0^2 \Omega_m$, with $\bar{\rho}_{\mathrm{m}0}$ being the background matter density today, and $\Omega_m$ the present-day fractional energy density of matter. $H_0$ is the Hubble expansion rate today. $c_1, c_2$ and $n$ are free parameters of the theory. One can then show that:

\bq \label{minHuS}
f_R = -\frac{c_1}{c_2^2} \frac{n\left( -R/M^2\right)^{n-1}}{\left[ \left(-R/M^2 \right)^n + 1  \right]^2}\;.
\eq

Given that:

\bq
- \bar{R} \approx 8 \pi G \bar{\rho}_m - 2 \bar{f}(R) = 3M^2 \left[ a^{-3} + \frac{2}{3} \frac{c_1}{c_2} \right]\;,
\eq

in order to match the $\Lambda$CDM background expansion, we set $c_1/c_2 = 6 \Omega_\Lambda / \Omega_m$. In this paper, we use $\Omega_m = 0.281$ and $\Omega_\Lambda \equiv 1 - \Omega_m = 0.719$ from WMAP9 \cite{2013ApJS..208...19H}. In doing so, we find that $-\bar{R} \approx 34M^2 \gg M^2$, \tcr{so that we can further simplify Eq.~\ref{minHuS} as}:

\bq
f_R \approx -n \frac{c_1}{c_2^2} \left[ \frac{M^2}{-R} \right]^{n+1}\;.
\eq

Finally, we define $\xi \equiv c_1/c_2^2$, and essentially reduce the Hu-Sawicki model into a two-parameter family in $\left(n, \xi\right)$. \tcr{This is because once the background evolution is fixed to match that of $\Lambda$CDM as a good approximation, it is only the combination $c_1/c_2^2$ that appears in the $f(R)$ field equations.}

\section{$f(R)$ Equations}
\label{freqs}

\subsection{The \tcr{Newtonian-gauge Perturbation Variables}}
\label{gauge}

In what follows, we shall 
work in the Newtonian gauge, defined in Eq.~\ref{Ngauge}. With the usual definitions of the Christoffel coefficients and the Ricci tensor as:

\bq \label{Christoffel}
\Gamma^\gamma_{\alpha \beta} &=& \frac{1}{2}g^{\gamma \eta} \left(\partial_\beta\, g_{\alpha \eta} + \partial_\alpha \,g_{\beta \eta} - \partial_{\eta}\, g_{\alpha \beta}\right),\;\rm{and} \\
R_{\mu \nu} &=& \partial_\gamma \,\Gamma^\gamma_{\mu \nu} - \partial_\nu \,\Gamma^\gamma_{\mu \gamma} + \Gamma^\lambda_{\gamma \lambda}\Gamma^\gamma_{\mu \nu} - \Gamma^\lambda_{\gamma \mu} \Gamma^\gamma_{\lambda \nu}\;,
\eq
where $\partial_\alpha$ is the 
\tcr{partial} derivative \tcr{with respect to $x^\alpha$}, we find, \tcr{up to first order in perturbation variables $\Phi$ and $\Psi$},
\bq \label{GaugeEq}
\Gamma^0_{00} &\approx& \dot{\Psi}\;, \nonumber\\
\Gamma^0_{0i} &\approx& \partial_i\,\Psi\;, \nonumber\\ 
\Gamma^i_{00} &\approx& \frac{1}{a^2} \delta^{ij} \partial_j\,\Psi\;, \nonumber \\
\Gamma^i_{j0} &\approx& \left(H - \dot{\Phi}\right) \delta^i_j \;, \nonumber \\
\Gamma^0_{ij} &\approx& a^2 H \left(1 - 2\Psi -2\Phi \right)\delta_{ij} - a^2\dot{\Phi}\delta_{ij}\; ,\nonumber \\
\Gamma^i_{jk} &\approx& -\partial_k \, \Phi \delta^i_j - \partial_j \Phi \delta^i_k + \partial^i \Phi \delta_{jk}\;,
\eq
where the overdots indicate derivatives with respect to the \tcr{physical time $t$}, and $H = \dot{a}/a$. The corresponding Ricci tensor components are:
\bq \label{Ricci}
R_{00} &\approx& \frac{1}{a^2} \Psi^{,i}_{\,,i} - 3 \left(\dot{H} + H^2\right) \nonumber \\ &&+ 3\ddot{\Phi} + 3H\left(\dot{\Psi} + 2\dot{\Phi}\right)\;, \\
R_{0i} &\approx& 2\dot{\Phi}_{,i} + 2 H \Psi_{,i}\;, \\
R_{ij} &\approx& \left(\Phi - \Psi\right)_{,ij} + \Phi^{,k}_{\,,k} \delta_{ij} - a^2 \ddot{\Phi} \delta_{ij} \nonumber \\  &&+ a^2\left(\dot{H} + 3H^2\right)\left(1 - 2\Phi - 2\Psi\right)\delta_{ij} \nonumber \\ &&- a^2 H \left(\dot{\Psi} + 6\dot{\Phi}\right) \delta_{ij}\;.
\eq
By using the definition of the Ricci scalar:
\bq
R = g^{\mu \nu} R_{\mu \nu}\;, 
\eq
in conjunction with Eq.~\ref{Ngauge}, we obtain:

\bq \label{Riccisc}
R &\approx& \frac{1}{a^2} \left(2 \Psi^{,i}_{\, ,i} - 4 \Phi^{,i}_{\, ,i}\right) + 6\ddot{\Phi} + 6H\left(\dot{\Psi} + 4\dot{\Phi}\right) \nonumber \\ &&- 6\left(\dot{H} + 2H^2\right)\left(1-2\Psi\right)\;.
\eq
Finally, with the definition of the Einstein tensor as:
\bq
G^\mu_{\,\nu} = R^\mu_{\, \nu} - \frac{1}{2} \delta^\mu_{\, \nu} R\;,
\eq
we find:

\bq \label{Einstein}
G^0_{\,0} &\approx& \frac{2}{a^2} \Phi^{,i}_{\, ,i} + 3H^2 - 6H\left(\dot{\Phi} + H \Psi\right) \nonumber\;, \\
G^0_{\, i} &\approx& 2\dot{\Phi}_{\,, i} + 2H \Psi_{, i} \nonumber\;, \\ 
G_{ij} &\approx& \left(\Phi - \Psi\right)_{, ij} + \left(\Psi - \Phi\right)^{,k}_{\, ,k} \delta_{ij} + 3a^2 \ddot{\Phi} \delta_{ij} \nonumber \\
&& 2 a^2 H^2\left(1 - 2\Psi\right)\delta_{ij} - 3H^2 a^2 \left(1-2\Psi\right)\delta_{ij} \nonumber \\  &&+ a^2 H \left(2\dot{\Psi} + 6\dot{\Phi}\right) \delta_{ij}\;.
\eq

\subsection{The Modified $f(R)$ Equation of Motion}
\label{MEOM}

The scalaron equation of motion (Eq.~\ref{scEOM}) assumes the quasi-static approximation (i.e., the time derivatives of the scalaron field are neglected), and hence \tcr{need to be generalised for the study here}. We therefore re-derive the equation of motion in the Newtonian gauge using Eqs.~\ref{Christoffel}--\ref{Einstein}. Using the definition that $\Box\,f_R = g^{\mu \nu} \nabla_\mu \nabla_\nu f_R$, we find that in the Newtonian gauge:

\bq \label{BoxNewton}
\Box\,f_R &=& \left(1 - 2 \Psi\right) \ddot{f_R} - \frac{1}{a^2} f_{R\;,i}^{, i} \nonumber \\ && + \left[ 3H\left(1 - 2\Psi\right) - \dot{\Psi} - 2\dot{\Phi} \right] \dot{f_R}\;.
\eq
When deriving Eq.~\ref{BoxNewton}, we have retained terms involving $\Psi$ $\dot{\Phi}, \dot{\Psi}$, but neglect second-order terms such as $\Phi^{,i}\Phi_{, i}$ and $\Phi \dot{\Phi}$. In what follows, we also make use of the following relations:

\bq \label{Approxs}
&&\left|\Phi\right| \sim \left|\Psi\right| \ll 1, \ \ \ \ |f_R| \ll 1, \ \ \ \ |\dot{\Phi}| \sim |\dot{\Psi}| \ll H, \nonumber \\
&& |\ddot{\Phi}| \sim |\ddot{\Psi}| \sim H|\dot{\Psi}| \sim H|\dot{\Phi}| \ll H^2 \sim |\dot{H}|\;,
\eq
so that quantitates on the left-hand side of the inequalities can be neglected when compared to the terms on the right-hand sides.

Since we are interested in the effects \tcr{of the field perturbations}, we need to subtract the contribution of the background quantities 
from these equations. Denoting such quantities with an overbar, and using Eq.~\ref{FREOM}, 
we write the following background equation of motion for the scalaron:

\bq \label{background}
\ddot{\bar{f}}_R + 3H \dot{\bar{f}}_R &\approx& \frac{1}{3} \left[    \bar{R} - \bar{f}_R \bar{R} + 2\bar{f}_R (R) + 8 \pi G \bar{\rho}_m   \right] \nonumber \\
&\approx& 0,
\eq
where the second equality comes from the assumption that, at the background level, the scalaron field $\bar{f}_R$ always follows the minimum of its effective potential. In reality $\bar{f}_R$ oscillates quickly around the minimum because $m_{\rm eff}^2\gg H^2$, such that over many oscillations the above assumption describes the average effect well (we will revisit to this point below). Under this assumption, and because
the value of the scalaron itself is quite small ($| \bar{f}_R| \leq 10^{-4}$ in the models studies here), we can assume that $|\bar{f}_R \bar{R}| \ll |\bar{R}|$, and rewrite Eq.~\ref{background} as:

\bq\label{RRR}
- \bar{R} &=& 8 \pi G \bar{\rho}_m + 2 \bar{f}_R (R)  \nonumber \\
&\approx& 8 \pi G \bar{\rho}_m + 32 \pi G \bar{\rho}_\Lambda \nonumber \\
&=& 8 \pi G T^\mu_{\, \mu },
\eq
\tcr{where we have used the fact that when $|R|\gg M^2$, which always holds for the models studied here, $f(R)$ remains approximately constant throughout the cosmic history (cf.~Eq.~\ref{HuS}). Note that the fact that $f(R)$ remains approximately constant for different values of $R$ means also that its perturbations are small and can be neglected, namely}
\bq
\tcr{f(R)-\bar{f}(R)\ \sim\ f_RR-\bar{f}_R\bar{R}\ \ll\ R-\bar{R}.}
\eq
Subtracting off the background \tcr{part from the scalaron equation of motion}, and denoting the perturbed quantities as $R - \bar{R} \equiv \delta R$ and $\rho_m - \bar{\rho}_m \equiv \delta \rho_m$, \tcr{we find}:
\bq \label{EOM2}
\ddot{f}_R + 3H \dot{f}_R - \frac{1}{a^2} \vec{\nabla}^2 f_R \approx \frac{1}{3} \left[  \delta R + 8 \pi G \delta \rho_m \right].
\eq

Note that the use of Eq.~\ref{background} implicates that it is $\ddot{f_R}$ that appears in this equation, rather than $\delta \ddot{f_R}$. \tcr{This is convenient because later we will write $\delta R$ as a function of $f_R$ instead of $\delta f_R=f_R-\bar{f}_R$}.

A quick comparison to the quasi-static version of the $f(R)$ equation of motion (Eq.~\ref{scEOM}) shows that the first two terms on the  left-hand side of Eq.~\ref{EOM2} are the additional terms one is left with when keeping the time derivatives in the scalar field equation of motion.

\subsection{The Modified Poisson Equation}
\label{modPE}

Eq.~\ref{EOM2} is one of the two equations that govern the formation of structure -- the other is the modified Poisson equation. The full Einstein field equations in $f(R)$ gravity become:

\begin{widetext}
\bq
\left(1+f_R\right) G_{\mu \nu} &=& 8 \pi G T_{\mu \nu} + \left[ \frac{1}{2} f(R) - \frac{1}{2} f_R R - \Box f_R\right] g_{\mu \nu} + \nabla_\mu \nabla_\nu f_R\;,
\eq
\end{widetext}

with the \tcr{following individual space-time components written in the Newtonian gauge}:

\begin{widetext} 
\bq \label{Components}
\frac{2}{a^2} \Phi^{, i}_{\, ,i} + 3H^2 &\approx& \frac{16}{3} \pi G \rho_m - \frac{1}{3} R - \frac{1}{6} f(R) + \ddot{f}_R \ \ \ \ \text{($0-0$ component\tcr{, full}),} \nonumber \\
\frac{2}{a^2} \Phi^{, i}_{\, ,i} &\approx& \frac{16}{3} \pi G \delta \rho_m - \frac{1}{3} \delta R + \ddot{f}_R \ \ \ \ \text{($0-0$ component, excluding background)}, \nonumber \\
\frac{1}{a^2} \left(\Psi - \Phi\right)^{,i}_{\,,j} + \delta^i_j \left[  2\dot{H} + 3H^2 - \frac{1}{a^2}\left(\Psi - \Phi\right)^{,k}_{\, ,k} \right]
 &\approx& 8 \pi G T^i_{\, j} - \frac{8}{3} \pi G \rho_m \delta^i_j - \frac{1}{3} R \delta^i_j - \frac{1}{6} f(R) \delta^i_j - \frac{1}{a^2} f_{R\; ,j}^{,i} + \dot{f}_R H \delta^i_{\,,j} \nonumber \\
 && \qquad \qquad \qquad \qquad \qquad \qquad \qquad  \text{($i-j$ components\tcr{, full}),} \nonumber \\
\frac{2}{a^2}\left( \Psi - \Phi \right)^{,i}_{\,,i} - 9H^2 + 6\dot{H} &\approx& 8 \pi G \left(\rho_m + 3 p_m\right) + R  + \frac{1}{2}f(R) +\frac{1}{a^2} f_{R\; ,i}^{, i} - 3H \dot{f}_R \nonumber \\ && \qquad \qquad \qquad \text{(Trace of $i-j$ components, including background),} \nonumber \\
\frac{2}{a^2} \left( \Psi - \Phi\right)^{,i}_{\,,i} &\approx& 8 \pi G \left(\rho_m + 3p_m\right) - 8 \pi G \left(\bar{\rho}_m + 3 \bar{p}_m\right) + \delta R + \frac{1}{a^2} f_{R\;,i}^{,i} - 3H \dot{f}_R \nonumber\\ &&\qquad \qquad \qquad \text{(Trace of $i-j$ components, excluding background).}
\eq
\end{widetext}
\tcr{In the above, the equations marked as `excluding background' are obtained by directly subtracting the $\Lambda$CDM background Friedmann equations from the full (00) and ($ij$) components of the modified Einstein equations, and using $\bar{f}(R) \approx 16\pi G\rho_{\Lambda}$ (cf.~Eq.~\ref{RRR}). This is why terms such as $\ddot{f}_R$ and $3H\dot{f}_R$ appear in them, rather than $\ddot{f}_R-\ddot{\bar{f}}_R$ and $3H\dot{f}_R-3H\dot{\bar{f}}_R$.}

\tcr{The Poisson equation can be obtained by taking the trace of the Einstein field equation, and from this we get}:
\bq \label{ModPoisson}
\frac{1}{a^2} \vec{\nabla}^2 \Psi \approx \frac{16 \pi G}{3} \delta \rho_m + \frac{1}{6} \delta R + \ddot{f}_R\;,
\eq
where Eq.~\ref{EOM2} has been used to eliminate $\frac{1}{a^2} f_{R\, ,i}^{,i}- 3H\dot{f_R}$ in Eqs.~\ref{Components}.

Eq.~\ref{ModPoisson} alongside Eq.~\ref{EOM2} are the two that we need to solve and use to update the simulation particle positions to \tcr{quanfity} the effect of non-vanishing time derivatives of $f_R$.

\tcr{We would like to} make a final note before concluding this section. In principle, terms such as $H \dot{\Psi}$ can be of the order of $H \dot{f_R}$, even though we have neglected them here. In our investigation, however, the aim is not to numerically solve all possible non-static terms, but rather to consistently investigate the effects of terms in $\dot{f_R}$ and $\ddot{f_R}$. Therefore, even though our equations are in some sense incomplete, they are sufficient for our specific purpose here.

\section{Evolution Equations in ECOSMOG}
\label{ECO}

Our $N$-body \tcr{simulations are performed} using the massively-parallelised {\sc ecosmog} code \cite{2012JCAP...01..051L}, which is based on the adaptive mesh refinement (AMR) code {\sc ramses} \cite{2002A&A...385..337T}. An AMR code can resolve high-density regions by refining (i.e., splitting) a mesh cell into eight sub-cells, when the number of particles within it exceeds some predefined threshold. This is particularly useful in $f(R)$ gravity simulations, where it \tcr{is} necessary in the high-density regions to \tcr{achieve} adequate resolution in order to solve the non-linear field equations and accurately quantify the chameleon effect. The code \tcr{employs} a multigrid relaxation algorithm, \tcr{arranged in V-cycles}, to accelerate the convergence of the solution~\cite{2002nrc..book.....P}.
\\

\subsection{Equations in Code Units}
\label{codeUnits}

In order to solve Eq.~\ref{EOM2} and Eq.~\ref{ModPoisson}, we need to convert the quantities in those equations to the superconformal units used by {\sc ecosmog}, summarised in \tcr{the equations} below:

\bq \label{CodeUnits}
\tilde{x}\ =\ \frac{x}{B},\ \ \ \tilde{\rho}\ =\ \frac{\rho a^3}{\rho_c\Omega_m},\ \ \ \tilde{v}\ =\ \frac{av}{BH_0},\nonumber\\
\tilde{\Psi}\ =\ \frac{a^2\Psi}{(BH_0)^2},\ \ \ \mathrm{d}\tilde{t}\ =\ H_0\frac{\mathrm{d}t}{a^2},\ \ \ \tilde{c}\ =\ \frac{c}{BH_0}, \nonumber \\
\tilde{f}_R =\ a^2 f_R\;.
\eq

Here, $x$ is the comoving coordinate,  $a$ is the scale factor, $\rho_c$ the critical density of the Universe today, $v$ is the particle velocity, $\Psi$ is the gravitational potential and $c$ is the speed of light. Furthermore, $B$ is the comoving size of the simulation box in units of $h^{-1}$ Mpc, whereas $H_0 =100\,h$ kms$^{-1}$ Mpc$^{-1}$. Under these conventions, the new terms appearing from the inclusion of the time derivatives become:

\bq \label{fddot}
\dot{f}_R &=& a^{-2} \dot{\tilde{f}}_R - 2 a^{-2} H \tilde{f}_R \nonumber \\
\ddot{f}_R &=& a^{-2} \ddot{\tilde{f}}_R - 4a^{-2} H \dot{\tilde{f}}_R - 2a^{-2} \dot{H} \tilde{f}_R \nonumber \\ &&+ 4a^{-2} H^2 \tilde{f}_R\;.
\eq

For the Hu-Sawicki model, this then transforms the modified Poisson equation (Eq.~\ref{ModPoisson}) and the $f(R)$ equation \tcr{of motion} (Eq.~\ref{EOM2}) into:

\begin{widetext}
\bq \label{FullEqs}
\tilde{\nabla}^2 \tilde{\Psi} &=& 2 \Omega_m a \left(\tilde{\rho} - 1\right) + \frac{1}{6} \Omega_m a^4 \left[   \left(  \frac{- n a^2 \xi}{\tilde{f}_R} \right)^{\frac{1}{n+1}} - 3 \left(  a^{-3} + 4 \frac{\Omega_\Lambda}{\Omega_m} \right)  \right] + \left[  a^{-2} \frac{\mathrm{d}^2 \tilde{f}}{\mathrm{d}\tilde{t}^2} - 6 \frac{H}{H_0} \frac{\mathrm{d}\tilde{f}}{\mathrm{d} \tilde{t}} + 2a^2 \left(  2 \frac{H^2}{H_0^2} -  \frac{\dot{H}}{H_0^2} \right) \tilde{f}_R  \right]\;, \nonumber \\
\tilde{\nabla}^2 \tilde{f}_R &=& \frac{-1}{\tilde{c}^2} \Omega_m a \left(\tilde{\rho} -1\right) - \frac{1}{3\tilde{c}^2} \Omega_m a^4 \left[ \left(  \frac{- n a^2 \xi}{\tilde{f}_R} \right)^{\frac{1}{n+1}} - 3 \left(  a^{-3} + 4 \frac{\Omega_\Lambda}{\Omega_m} \right)   \right] \nonumber
\\&&+ \frac{1}{\tilde{c}^2} \left[ a^{-2} \frac{\mathrm{d}^2 \tilde{f}}{\mathrm{d}\tilde{t}^2} - 3 \frac{H}{H_0} \frac{\mathrm{d}\tilde{f}}{\mathrm{d} \tilde{t}} -2a^2 \left(  \frac{H^2}{H_0^2} + \frac{\dot{H}}{H_0^2} \right) \tilde{f}_R   \right]
\eq
\end{widetext}

Note that all terms in the above equations are dimensionless \tcr{(dimensional quantities, such as $H$ and $\dot{H}$, are properly normalised using $H_0$)}. We have also carefully distinguished between overdots (derivatives with respect to the physical time $t$) and $\mathrm{d}/\mathrm{d} \tilde{t}$ (derivatives with respect to the superconformal time $\tilde{t}$), such that the former only applies to purely background quantities such as $a$ and $H$. Since the background evolution is approximated in the same way as in $\Lambda$CDM, quantities such as $H/H_0$ and $\dot{H}/H_0^2$ can be obtained analytically.

\subsection{Discretising the Equations}
\label{Disc}

In this section, we discretise the equations in Eq.~\ref{FullEqs} to make them appropriate for implementation in {\sc ecosmog}. During its time evolution, the value of $\tilde{f}_R$ can be very close to zero, and to avoid numerical problems, we solve for a different variable, $\tilde{f}_R \equiv -e^{\tilde{u}}$, instead. The current value of a quantity $\phi$ in the grid cell $\left(i,j,k\right)$ will be identified as $\phi_{i,j,k}$. Since everything presented below is already in the code units, we will drop tilde symbols in the discretised Poisson and $f(R)$ equations for clarity \tcr{wherever this will not cause confusion (we keep the tilde in $\tilde{c}$, however)}. Given a cell size $h$, we obtain for the Poisson equation:
\begin{widetext}
\bq \label{discPoisson}
&&\frac{1}{h^2} \left[\Psi_{i+1,j,k} + \Psi_{i-1,j,k} + \Psi_{i,j+1,k} + \Psi_{i,j-1,k}+\Psi_{i,j,k+1}+\Psi_{i,j,k-1}-6\Psi_{i,j,k} \right] \nonumber\\&& = 2 \Omega_m a \left(\rho_{i,j,k} - 1\right) - \frac{1}{6} \Omega_m a^4 \left[ \left(n a^2 \xi\right)^{\frac{1}{n+1}} \exp\left({\frac{-u_{i,j,k}}{n+1}} \right)- 3 \left(a^{-3} + 4 \frac{\Omega_\Lambda}{\Omega_m}\right) \right] \nonumber \\
&& + \left[a^{-2} \Delta t^{-1} \left[\mathrm{d}f\mathrm{d}t_{i,j,k}^{(n)} - \mathrm{d}f\mathrm{d}t_{i,j,k}^{(n-1)} \right]
+ 6 \frac{H}{H_0} \Delta t^{-1} \exp\left(u_{i,j,k}\right)\left(u_{i,j,k} - u_{i,j,k}^{(n-1)}\right) - 2a^2\left(2 \frac{H^2}{H_0^2} - \frac{\dot{H}}{H_0^2} \right) \exp \left(u_{i,j,k}\right) \right]\;, \nonumber\\
\eq
\end{widetext}
where $\Delta t$ is the time step in code units adopted by the simulation. The last line in Eq.~\ref{discPoisson} contains the additional terms that arise from going beyond the quasi-static approximation. 

Discretising the $f(R)$ equation of motion is a similar, if slightly more laborious task. In order to reduce clutter, we define a variable $b \equiv e^u$\tcr{, and write the discrete scalaron equation as}

\begin{widetext}
\bq \label{discfR}
&&\frac{1}{h^2}\left[b_{i+\frac{1}{2},j,k}u_{i+1,j,k}-u_{i,j,k}\left(b_{i+\frac{1}{2},j,k}+b_{i-\frac{1}{2},j,k}\right)+b_{i-\frac{1}{2},j,k}u_{i-1,j,k}\right]\nonumber\\
&&+\frac{1}{h^2}\left[b_{i,j+\frac{1}{2},k}u_{i,j+1,k}-u_{i,j,k}\left(b_{i,j+\frac{1}{2},k}+b_{i,j-\frac{1}{2},k}\right)+b_{i,j-\frac{1}{2},k}u_{i,j-1,k}\right]\nonumber\\
&&+\frac{1}{h^2}\left[b_{i,j,k+\frac{1}{2}}u_{i,j,k+1}-u_{i,j,k}\left(b_{i,j,k+\frac{1}{2}}+b_{i,j,k-\frac{1}{2}}\right)+b_{i,j,k-\frac{1}{2}}u_{i,j,k-1}\right]\nonumber\\
&&+\frac{1}{3\tilde{c}^2}\Omega_ma^4\left(na^2\xi\right)^{\frac{1}{n+1}}\exp\left(-\frac{u_{i,j,k}}{n+1}\right) - \frac{1}{\tilde{c}^2}\Omega_ma\left(\delta_{i,j,k}-1\right)-\frac{1}{\tilde{c}^2}\Omega_ma^4\left(a^{-3}+4\frac{\Omega_\Lambda}{\Omega_m}\right)\nonumber\\
&&+\frac{1}{\tilde{c}^2} \left[- a^{-2} \Delta t^{-1} \mathrm{d}f\mathrm{d}t_{i,j,k}^{(n-1)} - a^{-2} \Delta t^{-2} \exp \left( u_{i,j,k}\right) \left(u_{i,j,k} - u_{i,j,k}^{(n-1)} \right) \right. \nonumber \\
&& \left. + 3 \frac{H}{H_0} \Delta t^{-1} \exp \left( u_{i,j,k}\right) \left(u_{i,j,k}^{(n)} - u_{i,j,k}^{(n-1)} \right) + 2a^2 \left( \frac{\dot{H}}{H_0^2} + \frac{H^2}{H_0^2} \right) \exp \left(u_{i,j,k} \right) \right] \ = \ 0.
\eq
\end{widetext}
Once again, the effect of the time derivatives is incorporated in the terms in the last two lines of Eq.~\ref{discfR}. 

We have seen \tcr{in Eqs.~\ref{FullEqs} that 
their discrete versions will contain} terms like $\mathrm{d}^2 \tilde{f} / \mathrm{d}\tilde{t}^2$ and $\mathrm{d}\tilde{f}/\mathrm{d} \tilde{t}$ in our code units. By discretising also in time, we find that:

\bq \label{discdfdt}
\frac{\mathrm{d}\tilde{f}}{\mathrm{d} \tilde{t}} &=& - \Delta t^{-1} \exp \left( u_{i,j,k}\right) \left(u_{i,j,k}^{(n)} - u_{i,j,k}^{(n-1)} \right)\nonumber\\
&\tcr{\equiv}& \tcr{{\rm d}f{\rm d}t^{(n)}_{i,j,k}},\nonumber\\
\tcr{\frac{\mathrm{d}^2\tilde{f}}{\mathrm{d} t^2}} &\tcr{=}&  \tcr{\Delta t^{-1}\left[{\rm d}f{\rm d}t^{(n)}_{i,j,k} - {\rm d}f{\rm d}t^{(n-1)}_{i,j,k}\right]},
\eq
in which $u_{i,j,k}^{(n)}$ and $u_{i,j,k}^{(n-1)}$ are respectively the values of the scalaron field in the \tcr{current} time step $(n)$ and the previous time \tcr{step  $(n-1)$}. Throughout this paper $u_{i,j,k}$ without a superscript $^{(n)}$ always denotes the value at step $(n)$. In the simulations, the code records $u^{(n)}_{i,j,k}$ and ${\rm d}f{\rm d}t^{(n)}_{i,j,k}$ for each cell so that in the step that follows, they can be used as $u^{(n-1)}_{i,j,k}$ and ${\rm d}f{\rm d}t^{(n-1)}_{i,j,k}$. Note that in principle we also need the value $u^{(n-2)}_{i,j,k}$ to evaluate ${\rm d}^2f/{\rm d}t^2$ at step $(n)$, but in practice this is implicitly included in the calculation of ${\rm d}f{\rm d}t^{(n-1)}_{i,j,k}$ at step $(n-1)$. 

\tcr{By doing the above, we are incorporating the time derivatives in an {\it implicit} way, in contrast to the {\it explicit} method that tries to evolve the scalar field by by 
\bq\label{explicit}
u^{(n)}_{i,j,k} = u^{(n-1)}_{i,j,k}+\frac{{\rm d}}{{\rm d}t}u^{(n-1)}_{i,j,k}\Delta t.
\eq
It is known that the implicit scheme of numerical integration is usually more stable than the explicit method. However, the main advantage of our method is that it does not change the property that the $f(R)$ equation, Eq.~\ref{discfR}, is a boundary-value problem and therefore can be solved using a relaxation algorithm, with very little change to the code structure of {\sc ecosmog}. The explicit scheme described in Eq.~\ref{explicit}, on the other hand, means that the equation becomes an initial-value problem. Of course, because we are evaluating the time derivatives in a `backward' manner (that is, we are computing ${\rm d}u^{(n)}/{\rm d}t$ and ${\rm d}^2u^{(n)}/{\rm d}t^2$ by using $u^{(n-1)}$ and $u^{(n-2)}$ rather than using variables evaluated at step $(n)$), this will inevitably introduce numerical errors in evolving the differential equation. However, by making the time steps short enough, the two methods should agree, and therefore a consistency check can always be done by reducing $\Delta t$ to 
confirm that the method works properly, as we will demonstrate below.}

\tcr{Another} important point needs to be made at this \tcr{stage}. \tcr{As mentioned above}, the value of the scalar field, $f_R$, oscillates quickly around the local potential minimum. \tcr{Therefore,} in order to calculate its time evolution, our procedure in Eq.~\ref{discdfdt} implicitly performs an average over the many oscillations 
in each time step of the simulation. Evaluating a more ``instantaneous'' time derivative \tcr{accurately} would require \tcr{a huge} number of time steps, 
\tcr{especially in high-density regions, where the scalaron mass $m_{\rm eff}$ is larger and so the scalar field oscillates faster, and is therefore not practical for our $f(R)$ simulations}.  
\tcr{For linear terms, such as $\ddot{f}_R$ and $H\dot{f}_R$, the order of doing the time average and solving the scalaron equation can be freely swapped, and therefore the procedure in Eq.~\ref{discdfdt} is expected to work without any problem. On the other hand, for the nonlinear terms in the scalaron equation, such as $\delta R(f_R)$, the order does matter, and using time-averaged values for $f_R$ will introduce errors which are expected to become larger if the nonlinearity gets stronger}. 
For our simulations, however, we do not expect such errors to be significant enough to affect our conclusions; we will revisit and quantify this point in Section~\ref{avgerrors}.  \\ 

Eq.~\ref{discfR} can be thought of \tcr{symbolically} as an equation involving a non-linear differential operator, in the form:
\bq
\mathcal{L}^h u_{i,j,k} = 0,
\eq
where the superscript $h$ indicates that the operator is acting on a level where the cell size is $h$. The Gauss-Seidel relaxation in {\sc ecosmog} then updates the scalar field as:
\bq \label{update}
u_{i,j,k}^{h,\,(\mathrm{new})} = u_{i,j,k}^{h,\,(\mathrm{old})} - \frac{\mathcal{L}^h \left(  u_{i,j,k}^{h,\,(\mathrm{old})} \right)}{\frac{\partial \mathcal{L}^h \left(  u_{i,j,k}^{h,\,(\mathrm{old})} \right)}{\partial u_{i,j,k}^{h,\,(\mathrm{old})}}}\;. 
\eq

The form of the denominator in the above equation is given by:
\begin{widetext}
\bq
&&\frac{\partial\mathcal{L}^h(u_{i,j,k})}{\partial u_{i,j,k}}\nonumber\\
&=& \frac{\tilde{c}^2}{2h^2}b_{i,j,k}\left[u_{i+1,j,k}+u_{i-1,j,k}+u_{i,j+1,k}+u_{i,j-1,k}+u_{i,j,k+1}+u_{i,j,k-1}-6u_{i,j,k}\right]\nonumber\\
&&-\frac{\tilde{c}^2}{2h^2}\left[b_{i+1,j,k}+b_{i-1,j,k}+b_{i,j+1,k}+b_{i,j-1,k}+b_{i,j,k+1}+b_{i,j,k-1}+6b_{i,j,k}\right]\nonumber\\
&&-\frac{1}{3(n+1)}\Omega_ma^4\left(na^2\xi\right)^{\frac{1}{n+1}}\exp\left(-\frac{u_{i,j,k}}{n+1}\right) \nonumber \\
&&+\frac{1}{\tilde{c}^2} \left[ \left( 3 \frac{H}{H_0} \Delta t^{-1} - a^{-2} \Delta t^{-2} \right) \exp \left(u_{i,j,k}\right) \left(1 + u_{i,j,k} - u_{i,j,k}^{(n-1)} \right) + 2a^2 \left( \frac{\dot{H}}{H_0^2} + \frac{H^2}{H_0^2}\right) \exp \left(u_{i,j,k}\right) \right]\;.
\eq
\end{widetext}
Again, in the above equation, the last line represents the additional terms that arise from the inclusion of the time derivatives, while the first \tcr{three} lines are exactly the same as in the ordinary quasi-static case.

\tcr{For more details about how the above discrete equations are implemented in {\sc ecosmog} and the associated technical details, such as the boundary conditions, the interested readers are referred to the original {\sc ecosmog} code paper \cite{2012JCAP...01..051L}.}

\subsection{Time Integration}

Since the main goal of our paper is to assess the importance of time derivatives in $f(R)$ simulations, the choice of time step is of fundamental importance. In {\sc ecosmog}, this is \tcr{determined using} the Courant-Friedrichs-Lewy (CFL) condition \cite{1928MatAn.100...32C, 2012JCAP...01..051L}, which is required for the stability of numerical integrations. \tcr{In our simulations, this} condition essentially requires that the size of \tcr{a physical} time step $\mathrm{d} t$ has to be smaller than the time it takes for a particle to travel to an adjacent grid \tcr{cell}. Denoting the particle velocity by $v$, and the physical size of a cell in the grid as $\mathrm{d} x$, then \tcr{the CFL condition dictates that} in a particular time step these quantities are linked by:
\bq \label{CFL}
\frac{1}{v}\frac{\mathrm{d} x}{\mathrm{d} t} \geq \mathcal{O}(1)\;.
\eq
This condition must be satisfied at every time step for the solution to be stable. Using that $v \ll c$, in code units (Eq.~\ref{CodeUnits}), this condition translates to:
\bq \label{CFLcode}
\frac{\mathrm{d} \tilde{x}^2}{a^2 \tilde{c}^2 \mathrm{d} \tilde{t}^2} \ll 1\;.
\eq
Recall (Eq.~\ref{CodeUnits}) that $ \tilde{c} = c / BH_0$, where $B$ is the box size of the simulation. For a fixed box size, Eq.~\ref{CFLcode} then tells us that:
\bq
h^2 \ll a^2 \tilde{c}^2 \Delta t^2,
\eq
\tcr{where $h={\rm d}\tilde{x}$ and $\Delta t={\rm d}\tilde{t}$}. 

\tcr{The above equation} already hints at the answer to our question regarding the importance of time derivatives relative to spatial derivatives. \tcr{It tells us that the} manner in which the size of the time step \tcr{($\Delta t$)} is \tcr{set} is such that it is generally much larger (with a multiplicative factor of $a\tilde{c}$) than the size of the cell \tcr{($h$)}. \tcr{As a result, in the discrete scalaron equation above (Eq.~(\ref{discfR}))}, one would expect the spatial variation of the scalar field \tcr{(terms proportional to $h^{-2}$)} to be more significant than its variation in time \tcr{(terms proportional to $(a\tilde{c}\Delta t)^{-2}$)} -- or, in other words, that a quasi-static approximation is good. In the following section, we will proceed to perform $N$-body simulations to confirm our expectation from these simple order-of-magnitude arguments.

\section{Results}
\label{Results}

In this section, we apply our modified {\sc ecosmog} code to perform $N$-body simulations of the Hu-Sawicki model, for three different choices of the present-day value of $\bar{f}_{R}$, 
namely $| \bar{f}_{R0}| = 10^{-4}, 10^{-5}, 10^{-6}$, which we will refer to as F4, F5 and F6 respectively. F4 (F6) forms an upper (lower) bound for cosmologically interesting $f(R)$ models: for $\left| \bar{f}_{R0} \right| > 10^{-4}$, the models are unlikely to satisfylocal gravity constraints in the Mily Way \cite{2009PhRvD..80h3505S}, whereas for $\left|\bar{f}_{R0}\right| < 10^{-6}$, the differences from GR are very small. In what follows, we also set the parameter $n = 1$ (Eq.~\ref{HuS}). 

The cosmological parameters for our $N$-body simulations are \tcr{the same as in the best-fitting WMAP9 cosmology \cite{2013ApJS..208...19H}, with} $\Omega_m = 0.281, \Omega_\Lambda = 0.719, h = \tcr{0.697}, n_s = 0.972$ and $\sigma_8 = 0.82$. Here, $h\tcr{~= H_0/(100~{\rm km/s/Mpc})}$ is the dimensionless Hubble parameter, $n_s$ is the spectral index of the primordial power spectrum, and $\sigma_8$ is the linear rms density fluctuation in a sphere of radius $8\,h^{-1}$ Mpc, at $z = 0$. We expect that our findings here should not change with a different choice of parameters.

We perform low-resolution runs for $\Lambda$CDM, F4, F5 and F6 in a box of size $256\,h^{-1}$ Mpc with $256^3$ particles, and higher-resolution runs for $\Lambda$CDM and F6 in a box size of $128\,h^{-1}$ Mpc with $256^3$ particles. In each case, we simulate 5 realisations of initial conditions \tcr{(the same initial conditions are used for $\Lambda$CDM and $f(R)$ simulations because at the initial time, $z_{i}=49$, the effect of modified gravity is still negligible for F6, F5 and F4)}. \tcr{For every $f(R)$ simulation we have performed for this work, we conduct both a quasi-static run and a time derivative run, to quantify the impact of including non-static effects}. To check for the influence of changing the size of the time step and of resolution, we simulate two additional models: $L_{256/2}$ and $L_{64}$. The former has the same parameters as the $L_{256}$ run, but here we artificially halve the time step that the {\sc ecosmog} code would naturally \tcr{adopt}. \tcr{The latter} constitutes our highest 
resolution run, \tcr{with} $256^3$ particles within a box of $64\,h^{-1}$ Mpc. In each set of simulations, \tcr{the regular simulation mesh has $256$ cells on each side, and} is adaptively refined when the number of particles within a cell is greater than 8. 
A summary of the simulation details is given in Table~\ref{tab:SimDet}.

\begin{table}[t]
\begin{tabular}{@{}l c c c r} 
\hline
\hline
Name &   Model &  $L_{\mathrm{box}}$  &  Particles &  Realisations \\ 
\hline 
$L_{256}$ & $\Lambda$CDM, F4, F5, F6 &  $256\,h^{-1}$ Mpc & $256^3$ & 5 \\
$L_{256/2}$ & $\Lambda$CDM, F6 & $256\,h^{-1}$ Mpc & $256^3$ & 1 \\
$L_{128}$ &  $\Lambda$CDM, F6 & $128\,h^{-1}$ Mpc & $256^3$ & 5 \\ 
$L_{64}$ & $\Lambda$CDM, F6 & $64\,h^{-1}$ Mpc & $256^3$ & 1 \\
\hline
\end{tabular}
\caption{Summary of simulations performed in this work.}
\label{tab:SimDet}
\end{table}

\subsection{The Matter and Velocity Divergence Power Spectra}
\label{sec:PS}

As remarked on earlier, the first order difference between \tcr{$f(R)$ simulations in the quasi-static approximation and the non-static limit can be seen in changes to the matter power spectrum \cite{2013PhRvD..88j3507H}.} In $f(R)$ gravity, one would expect the scalaron field\tcr{, through the fifth force it mediates (where the chameleon screening is not effective)}, to enhance the ordinary gravitational \tcr{interaction}, thereby \tcr{strengthening} the clustering of \tcr{matter}. To quantify this further, we define the dark matter density field $\rho \left(\vec{x}, t\right)$ as:
\bq 
\rho\left(\vec{x}, t\right) = \tcr{\bar{\rho}(t)[1+\delta(\vec{x},t)]},
\eq
where $\bar{\rho}$ is \tcr{the background density field at time $t$}, and $\delta$ encodes the fluctuations around that \tcr{homogeneous background}. In order to calculate the power spectrum, it is first convenient to rewrite the density contrast $\delta$ in Fourier space:
\bq
\delta_{\vec{k}} \equiv \left( 2\pi \right)^{-3/2} \int \delta \tcr{\left( \vec{x}, t \right)} e^{- i \vec{k} . \vec{x}} \mathrm{d}^3 \vec{x} \;.
\eq
The matter power spectrum is then defined by:
\bq \label{PS}
P_{\delta \delta} \left(k\right) \equiv P \left(k\right) = \left\langle \left| \delta_{\vec{k}} \right|^2 \right\rangle.
\eq
\tcr{To measure the matter power spectrum from our simulation outputs, we make use of the publicly-available {\sc powmes} code \cite{2011ascl.soft10017C}, which constructs the density field of a particle distribution by estimating the Fourier modes of the distribution using a Taylor expansion of trigonometric functions.}
We also compute the velocity divergence power spectra from our simulations, following the approach in~\cite{2013MNRAS.428..743L}. First, we define the {\it expansion scalar}, which is \tcr{related to the divergence of the velocity field by}:
\bq
\theta\tcr{\left(\vec{x}, t\right)} = \frac{1}{aH(a)} \vec{\nabla} \cdot v\tcr{\left(\vec{x},t \right)}\;,
\eq
where $v\tcr{\left(x, t\right)}$ is the cosmic peculiar velocity field and $H(a)$ is the Hubble constant at epoch $a$. In a similar vein to the matter power spectrum, we can take the Fourier transform of the above to get:
\bq 
\theta_{\vec{k}} \equiv \left( 2\pi \right)^{-3/2} \int \theta \tcr{\left(\vec{x}, t\right)} e^{- i \vec{k} . \vec{x}} \mathrm{d}^3 \vec{x} \;,
\eq
and the corresponding velocity divergence power spectrum:
\bq
P_{\theta \theta} \left(k\right) = \left< \left| \theta_{\vec{k}} \right|^2 \right>\;.
\eq

\begin{figure*}
\centering
\includegraphics[trim = 0mm 0mm 0mm 0mm, clip,width=0.8\linewidth, height = 0.6\textheight]{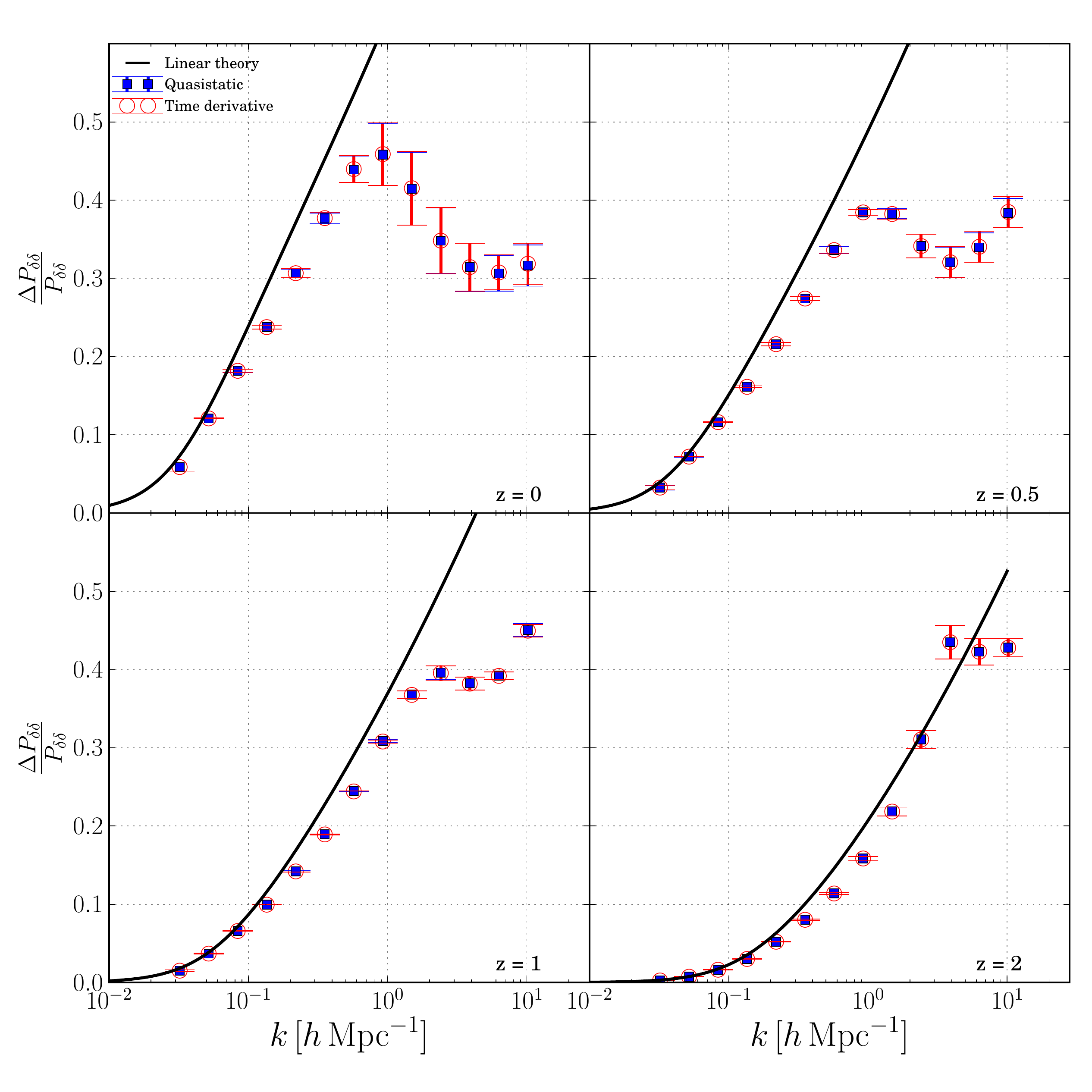}
\caption{Time evolution of excess clustering signal $\Delta P / P$ for the lower-resolution ($L_{256}$) F4 simulations, over four different redshifts. The open red circles represent the realisation-averaged relative difference when including the time derivatives, whereas the filled blue circles show the enhancement with respect to $\Lambda$CDM for the standard quasi-static case. The solid black line is the enhancement for quasi-static F4, relative to $\Lambda$CDM as predicted by linear theory. The procedure for calculating the averages and error bars is as described in the main text.}
\label{F4lowRes}
\end{figure*} 

The velocity field has been shown to be more sensitive than the matter \tcr{field} to the effects of the fifth force, so any changes due to the inclusion of time derivatives should also have a stronger signal here \cite{2011MNRAS.410.2081J}. We measure \tcr{$\theta(\vec{x}, t)$} from our simulation \tcr{outputs} by performing a Delaunay tessellation over the discrete set of points defining the configuration of our simulation, using the publicly available {\sc dtfe} code \cite{2000A&A...363L..29S, 2011arXiv1105.0370C}. This has the advantage of calculating a volume-weighted velocity divergence field, rather than a mass-weighted one, and also circumvents the issue of empty grid cells.

\begin{figure*}
\centering
\includegraphics[trim = 0mm 0mm 0mm 0mm, clip,width=0.8\linewidth, height = 0.6\textheight]{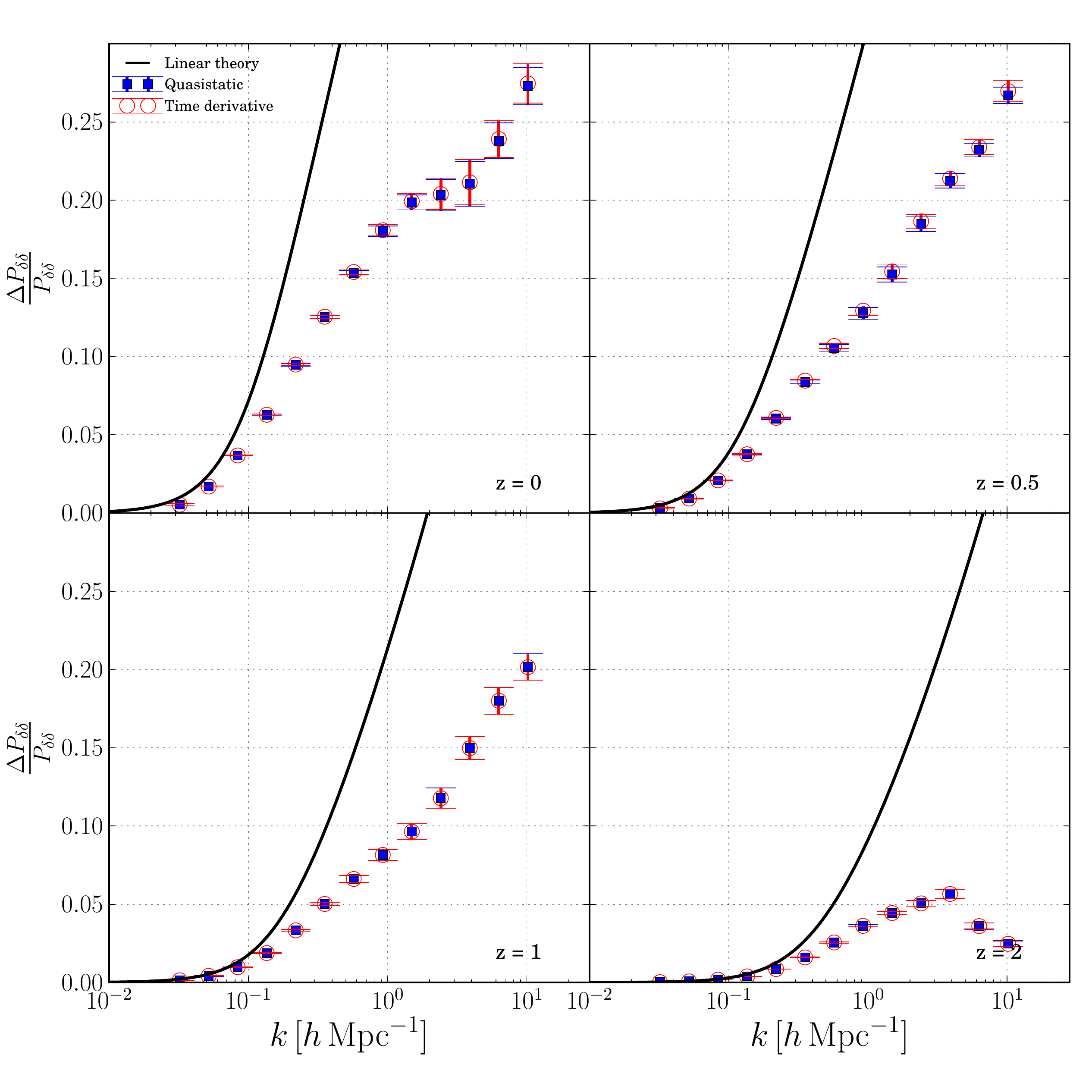}
\caption{Time evolution of $\Delta P / P$ for the lower-resolution ($L_{256}$) F5 simulations, over four different redshifts. The line/symbol styles are as described in Fig.~\ref{F4lowRes}.}
\label{F5lowRes}
\end{figure*}

\subsubsection{Low-resolution Tests}
\label{lowRes}

As a first test, we perform simulations with $256^3$ particles in the $L_{256}$ box. To see the difference between the simulation with time derivatives and that in the quasi-static limit, we measure the enhancement of the power spectrum in each case relative to $\Lambda$CDM. In what follows, we refer to the individual cases using the notation F$x_{\left\{q,t\right\}}$, where $x = 4, 5, 6$ \tcr{indicates} the value of $|\bar{f}_{R0}|$, while $q$ ($t$) refers to the simulation in the quasi-static limit (with the inclusion of time derivatives). 

We then smooth out the intrinsic noise in the power spectrum as follows. First, we calculate the relative difference in the power spectrum of F$x_{\left\{q,t\right\}}$ \tcr{compared with} $\Lambda$CDM in each set of realisations:
\bq \label{relDIff}
\frac{\Delta P(k; \mathrm{F}x_{\left\{q,t\right\}})}{P (k; \Lambda \mathrm{CDM})} = \frac{P(k; \mathrm{F}x_{\left\{q,t\right\}}) - P (k; \Lambda \mathrm{CDM})}{P (k; \Lambda \mathrm{CDM})}.
\eq
\tcr{We then divide the values of the wavenumber $k$ probed by the simulation into a number of bins equally spaced in $\log(k)$, and} average the relative difference in each bin over all the realisations. The scatter between realisations is represented by error bars calculated using the standard deviation in each $k$-bin over all realisations. The relative difference is taken with respect to $\Lambda$CDM, rather than between the quasi-static and non-static runs themselves, because the residual from the latter is expected to be very small, and taking the ratios of these small differences can look larger than they intrinsically are on a plot.

The results of the above procedure in the cases for F$4_{\left\{q,t\right\}}$, F$5_{\left\{q,t\right\}}$ and F$6_{\left\{q,t\right\}}$ are shown\tcr{, respectively,} in Fig.~\ref{F4lowRes}, \ref{F5lowRes} and \ref{F6lowRes}. Focusing \tcr{first} on the quasi-static (blue symbols) \tcr{simulations} only, we note two features consistent in F4, F5 and F6:
 
\begin{figure*}
\centering
\includegraphics[trim = 0mm 0mm 0mm 0mm, clip,width=0.8\linewidth, height = 0.6\textheight]{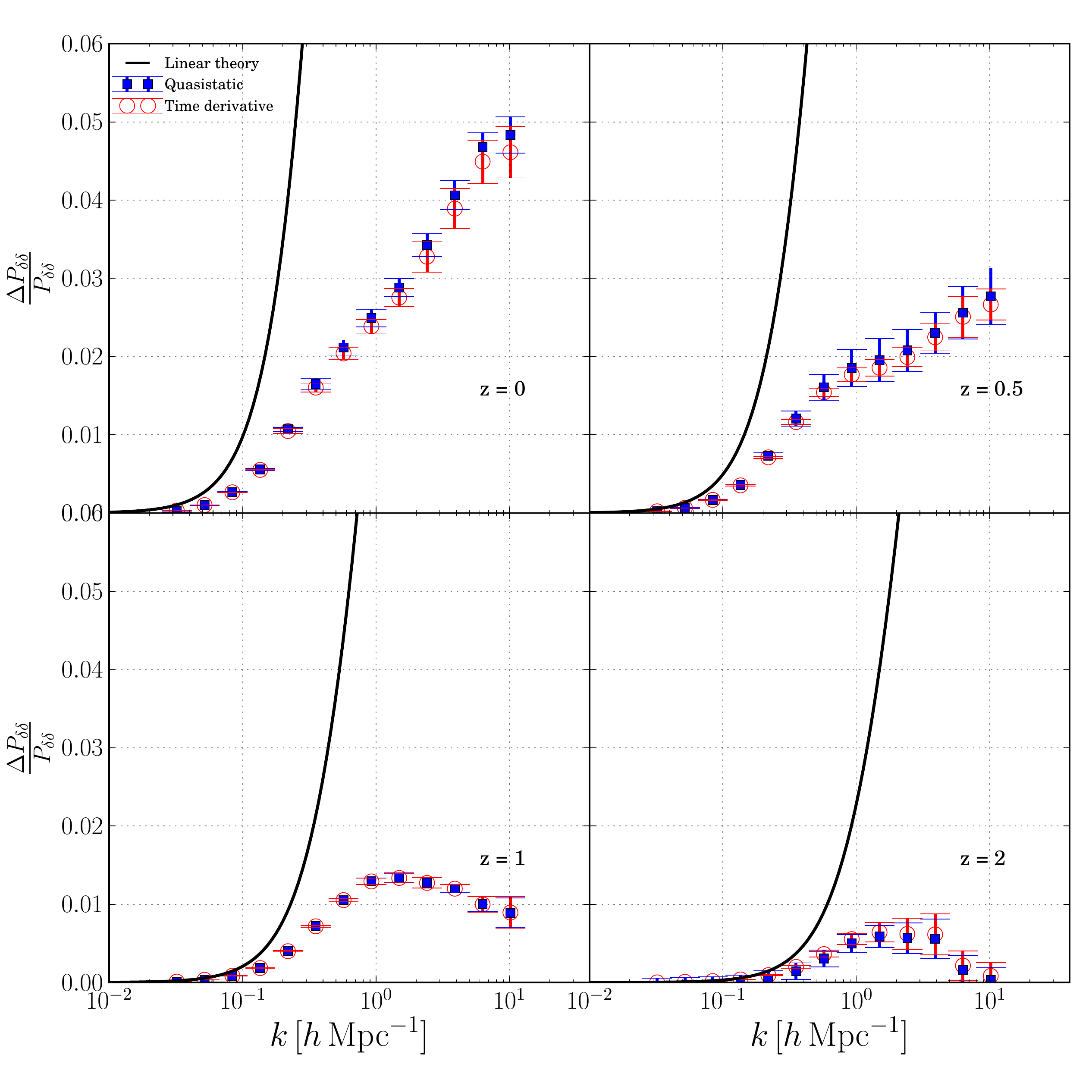}
\caption{Time evolution of $\Delta P / P$ for the lower-resolution ($L_{256}$) F6 simulations, over four different redshifts. Again, the line and symbol  styles follow the convention in Fig.~\ref{F4lowRes}.}
\label{F6lowRes}
\end{figure*}

Firstly, the enhancement in the \tcr{matter} power spectrum relative to $\Lambda$CDM closely follows the \tcr{predictions} of linear theory at large scales, which is what one would expect. At \tcr{smaller} scales, linear theory over-predicts the enhancement of power in the $f(R)$ model with respect to $\Lambda$CDM, because it fails to account for the suppression of the fifth force by the chameleon mechanism \tcr{and other nonlinear effects}. This can also be seen in Fig.~\ref{F4lowRes} for the F4 model, which is the one that most significantly deviates from GR -- it shows a better match to linear theory for $k \leq 1h\,\mathrm{Mpc}^{-1}$ compared to F5 and F6, because here the chameleon mechanism is less efficient.

Secondly, we have seen quite distinct features in $\Delta P_{\delta \delta} / P_{\delta \delta}$ for the three models. The amplitude of $\Delta P_{\delta \delta} / P _{\delta \delta}$ at $z = 0$ increases from F6 to F4, \tcr{which confirms} that the effect of the fifth force becomes stronger as the magnitude of the scalaron field $\left|\bar{f}_{R0}\right|$ increases. F4, for example, shows a distinct peak at around $k = 1\,h$Mpc$^{-1}$, as \tcr{demonstrated} in Fig.~\ref{F4lowRes}. In F5, at these scales $\Delta P_{\delta \delta} / P_{\delta \delta}$ shows a minor flattening before rising again to smaller scales. In F6, on the other hand, there are no such noticeable features, and the enhancement of the power spectrum increases all the way down to the smallest resolved scales. These features agree well with the results of \cite{2013MNRAS.435.2806H}, and can be explained by the different efficiency of the chameleon screening in the different models.

A look at these figures \tcr{leads us to} our main result, that there is no significant change in the clustering properties when we include time derivatives into our simulations. The differences, as can be gathered from the offset between the red and blue symbols, are sub-percent. If we now look at the effect of the time derivatives (open red circles),  we find that the smoothed results trace their quasi-static counterparts almost exactly. The error bars here, which represent the scatter in $\Delta P_{\delta \delta} / P_{\delta \delta}$ \tcr{across} realisations, 
\tcr{almost exactly overlap} as well. 
\tcr{This is particularly true for the F4 and F5 cases, as can be seen clearly from Figs.~\ref{F4lowRes} and \ref{F5lowRes}.}
\tcr{Towards} 
smaller scales, the discrepancy between the time derivative and quasi-static \tcr{runs} becomes slightly more pronounced, which is because the effects of time derivatives on the fifth force will be felt at the smallest scales first, due to the hierarchical nature of structure formation and the properties of the initial conditions.

\begin{figure}
\centering
\includegraphics[trim = 0mm 0mm 0mm 0mm, clip,width=\columnwidth, height = 0.45\textheight]{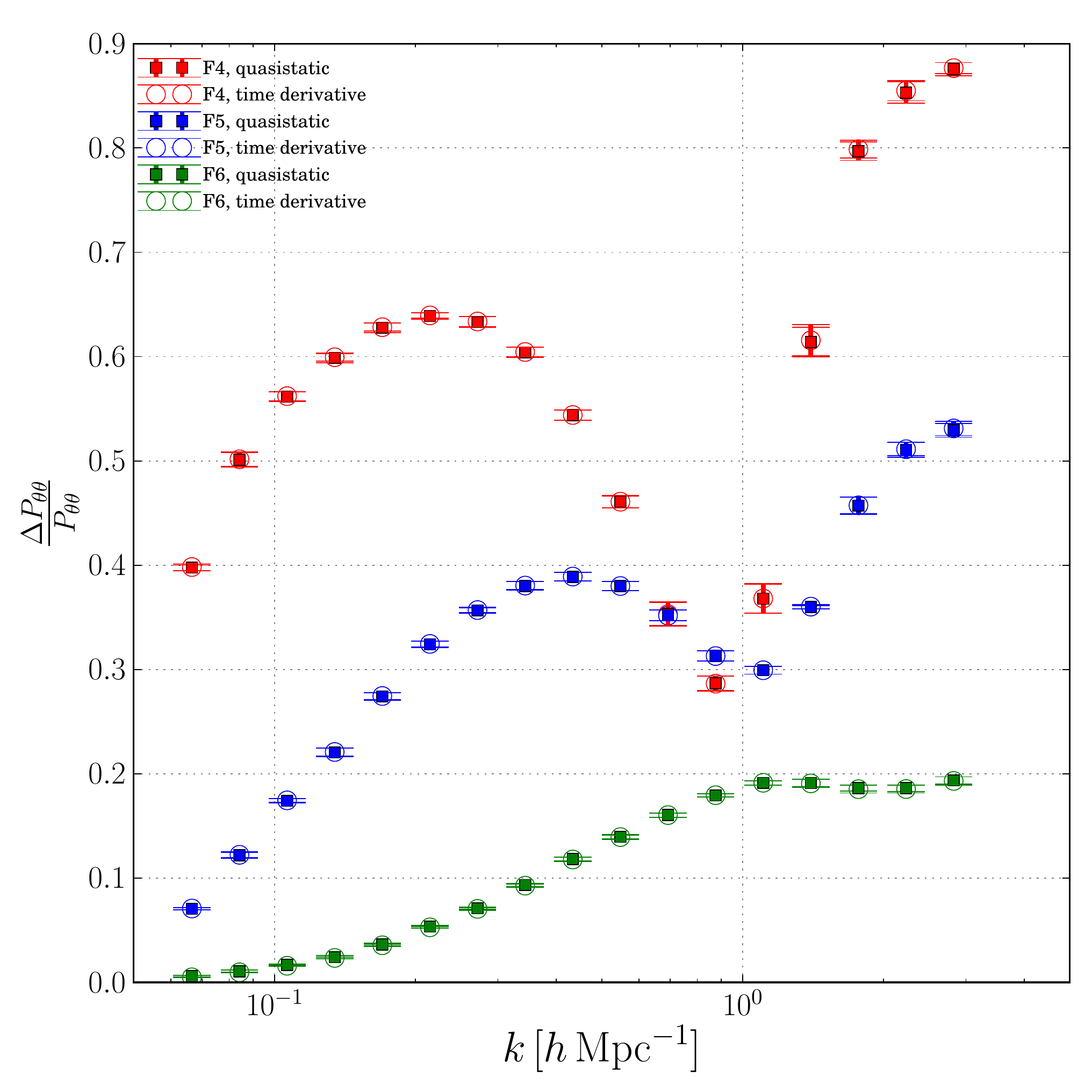}
\caption{The velocity divergence power spectrum at $z = 0$ for the F4, F5 and F6 models in the $L_{256}$ simulation.}
\label{velPS}
\end{figure}

Inspection of Fig.~\ref{F6lowRes} suggests that \tcr{the effect of time derivatives is more significant in F6} than in F4 or F5. \tcr{Here, a noticeable} offset between the \tcr{time derivative and quasi-static runs} starts as early as $z = 0.5$.  The larger effects of time derivatives 
\tcr{could be} because $\Delta P_{\delta \delta} / P_{\delta \delta}$ has a much smaller \tcr{magnitude} ($\leq5\%$ \tcr{down to $k\sim10~h$Mpc$^{-1}$}) in F6 than in F4 and F5, \tcr{which makes the small impact of including the time derivatives look much stronger}, \tcr{but it may also arise from numerical issues (e.g., the spatial and time resolutions of our simulations are too low and the results have not yet converged). To have confidence in using our numerical simulations to do science, it} is important then to understand whether this result is physical. 
For this reason, we need to investigate the differences between quasi-static and non-static runs when re-simulated at higher resolutions. We will return to this in the next subsection.

\begin{figure*}
\centering
\includegraphics[trim = 0mm 0mm 0mm 0mm, clip,width=0.8\linewidth, height = 0.6\textheight]{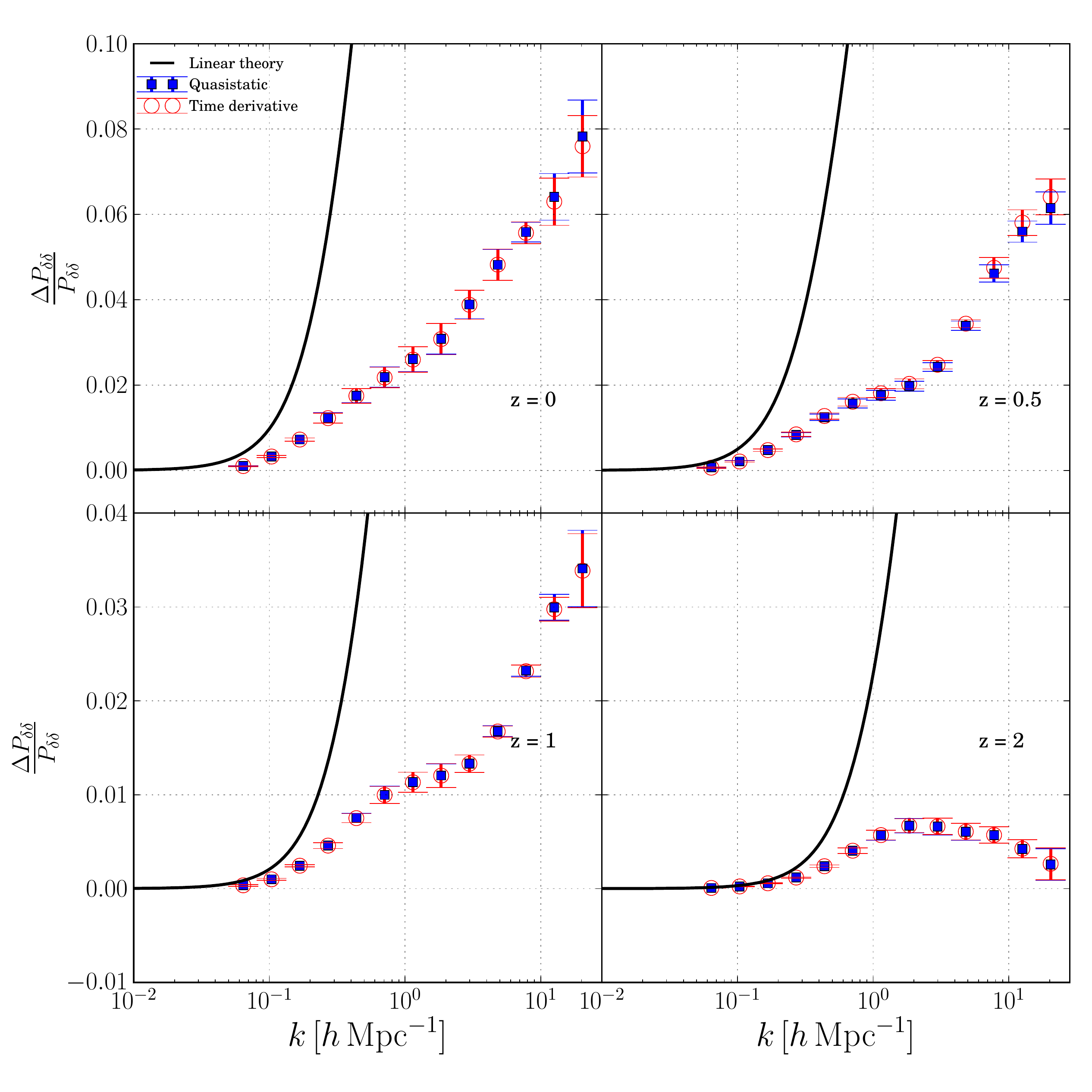}
\caption{Time evolution of $\Delta P / P$ for the higher-resolution ($L_{128}$) F6 simulations, over four different redshifts. Again, the line and symbol  styles follow the convention in Fig.~\ref{F4lowRes}.}
\label{F6highRes}
\end{figure*}

Finally, Fig.~\ref{velPS} illustrates the $z = 0$ relative difference in the velocity divergence power spectra ($\Delta P_{\theta \theta} / P_{\theta \theta}$) for \tcr{F4, F5 and F6}. All three models show similar features as first observed in \cite{2013MNRAS.435.2806H}, most markedly the presence of a dip, after which the ratio $\Delta P_{\theta \theta} / P_{\theta \theta}$ increases once again. Comparison with Figs.~\ref{F4lowRes}-\ref{F6lowRes} shows that the enhancement of $\Delta P_{\theta \theta} / P_{\theta \theta}$ for these models relative to $\Lambda$CDM is a lot stronger than that in the matter power spectra -- to almost an order of magnitude in the case of F6. This reiterates the aforementioned advantage of using the velocity divergence power spectrum as a more sensitive probe of modified gravity \tcr{\cite{{2014PhRvL.112v1102H}}. Just as in the case of the matter power spectrum, there does not seem to be any significant difference in the enhancements when including time derivatives, as both the non-
static and quasi-static \tcr{simulations} of the three models seem to be well-converged. \tcr{Note, however, that for F6 the effects of including time derivatives on $\Delta P_{\theta \theta} / P_{\theta \theta}$ appear to be much smaller than in the case of matter power spectra, which is because of the scale on the axis.}}

\subsubsection{High-resolution Tests}
\label{highRes}

\begin{figure*}
\centering
\includegraphics[trim = 0mm 0mm 0mm 0mm, clip,width=0.8\linewidth, height = 0.6\textheight]{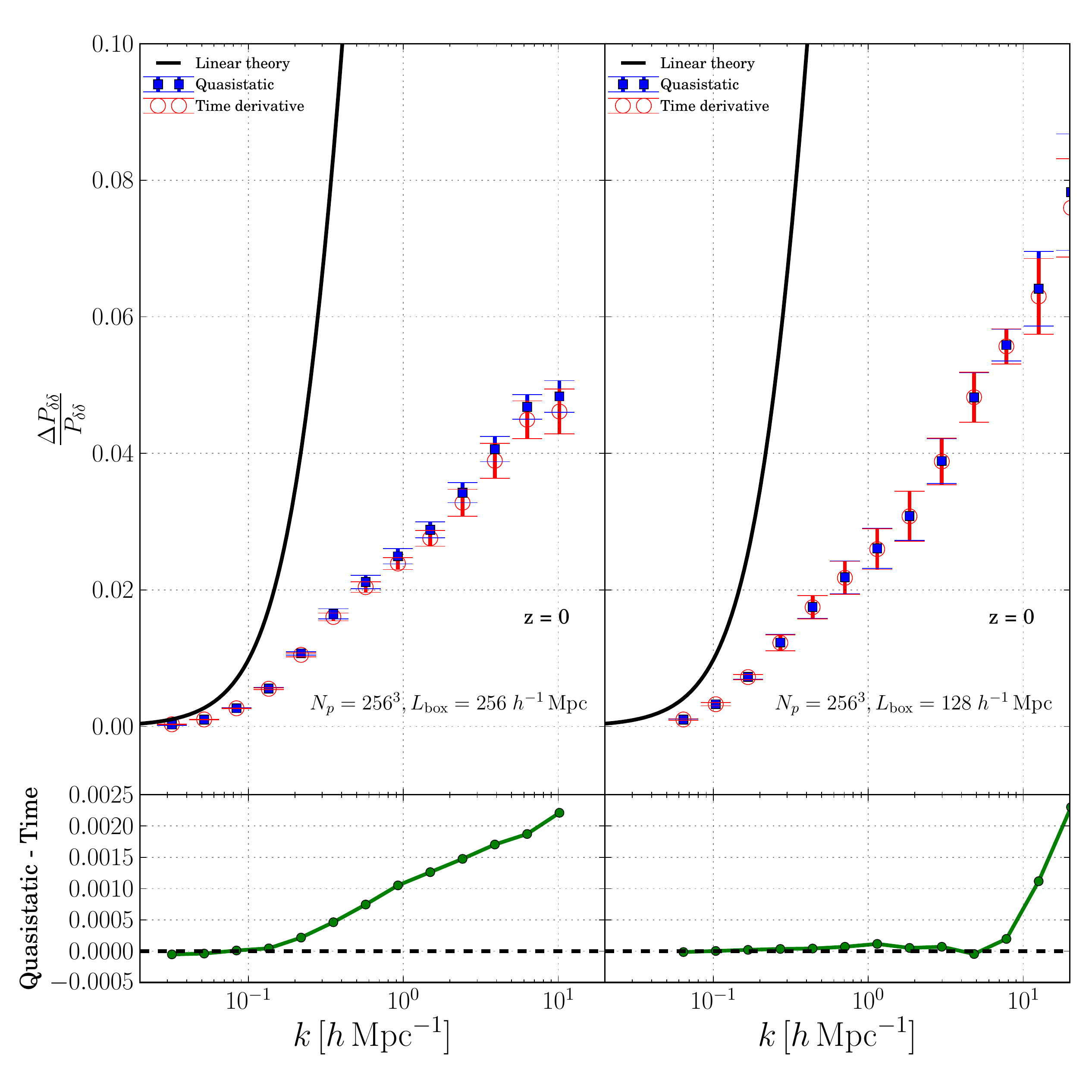}
\caption{(Left) $\Delta P / P$ for the lower-resolution ($L_{256}$) F6 simulations, shown at $z = 0$ in the top panel, with the relative difference between the quasi-static and non-static runs in the lower panel. There appears to be a systematic increase in the offset between the two cases with increasing $k$. We find that this discrepancy is purely a numerical artefact. (Right) The same is done for the $L_{128}$ simulation for F6 at $z = 0$. It is clear to see that increasing the resolution has led to a much improved convergence between the quasi-static and non-static cases (except at the very largest $k$).}
\label{F6ResDiff}
\end{figure*}

\begin{figure}
\centering
\includegraphics[trim = 0mm 0mm 0mm 0mm, clip,width=\columnwidth, height = 0.4\textheight]{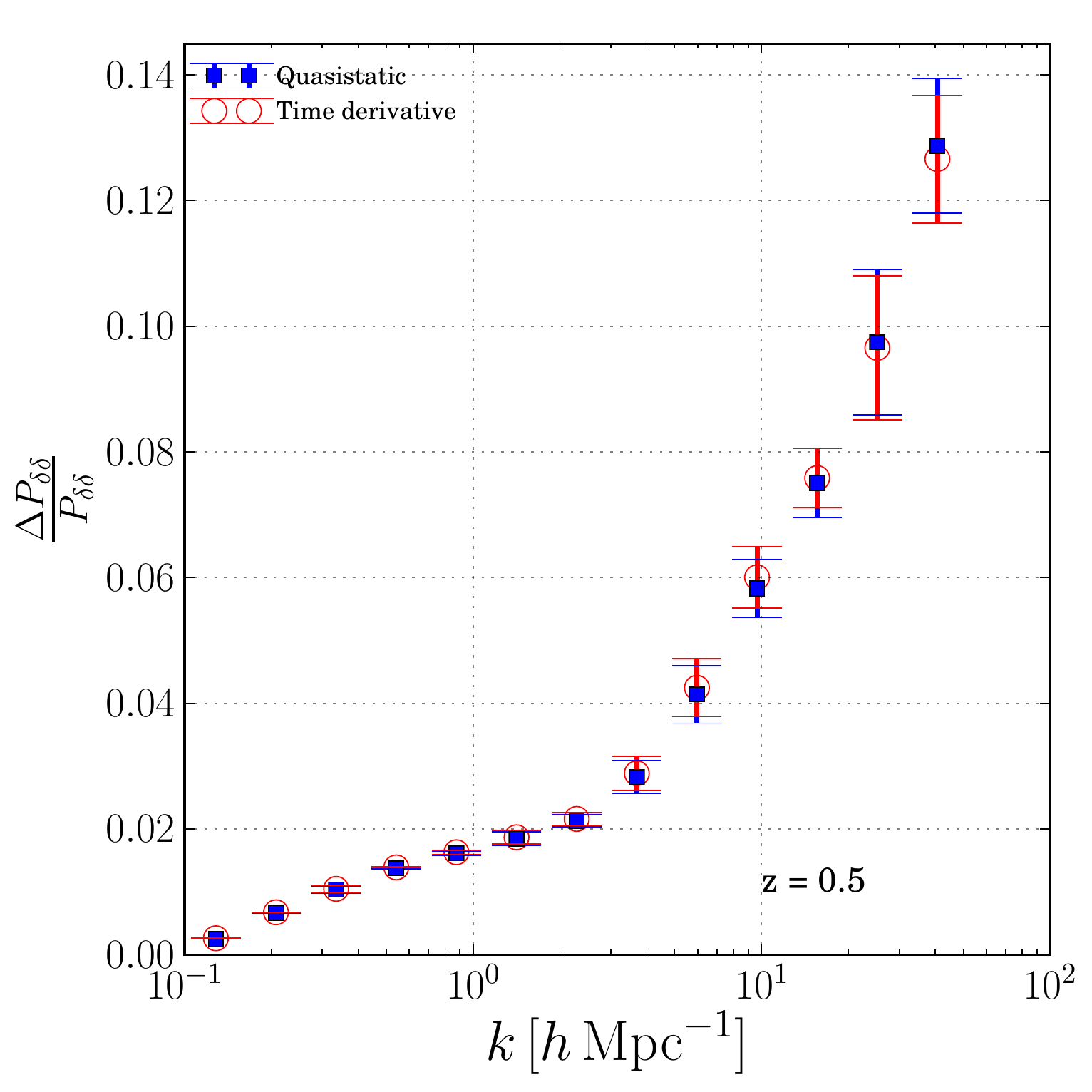}
\caption{$\Delta P / P$ for the highest-resolution ($L_{64}$) F6 simulations, at $z = 0.5$. Note that this was simulated using only one realisation, so the error bars represent the scatter in each bin of $\log (k)$. One can already see the marked improvement in the agreement between the quasi-static and time derivative simulations, compared to equivalent redshift and even $z = 0$ for the $L_{256}$ run (Fig.~\ref{F6lowRes}).}
\label{F6_64}
\end{figure}

We \tcr{have simulated} the F6 model at higher resolutions, by keeping the number of particles at $256^3$, \tcr{but using} smaller boxes of size $128\,h^{-1}$ Mpc and $64\,h^{-1}$ Mpc (the $L_{128}$ and $L_{64}$ simulations in our nomenclature). The result of the former is displayed in Fig.~\ref{F6highRes}, \tcr{from which 
we} immediately see that the discrepancy we noticed in Fig.~\ref{F6lowRes} is now largely reduced, even at redshift $z = 0.5$. This is demonstrated more clearly in Fig.~\ref{F6ResDiff}, where in the upper panels we again plot $\Delta P_{\delta \delta} / P_{\delta \delta}$ at $z = 0$ for both the $L_{256}$ and $L_{128}$ runs, and show the difference between the quasi-static and non-static cases for each in the lower panel. The offset seen earlier in the $L_{256}$ case is now essentially zero throughout all $k$ for $L_{128}$, except for the smallest scales (large $k$) where we are likely affected by resolution once more. \tcr{The case for the $L_{64}$ simulation is shown in Fig.~\ref{F6_64}, but only for the snapshot at $z=0.5$ (which shows the largest difference between the quasi-static and non-static  runs in Fig.~\ref{F6highRes}) for brevity. Again, here we see that the difference between the two is further reduced.}

\tcr{The implications of} the results shown in Figs.~\ref{F6highRes} and~\ref{F6_64} \tcr{are twofold. First, it serves as a convergence test of our algorithm to include time derivatives in the simulations and shows that, with increasing (spatial and time) resolution, the runs do converge as we anticipated. Second, it} resonates our expectations and findings from F4 and F5 models, that the effect of introducing time derivatives in the F6 model has a negligible impact on the matter power spectra, compared with just the quasi-static case \tcr{(if the resolution is high enough so that simulation has converged, of course)}. 

\tcr{Our conclusion is then that in all models studied in this work (which are also the most well-studied modified gravity models in the literature), the quasi-static approximation, which is adopted in almost all numerical simulations to date, is  valid and is adequate to make accurate predictions for the matter and velocity divergence power spectra.}


\subsection{Configuration Space}
\label{sec:clustamp_pdf}
So far we have focused on the quantities describing the cosmic density and velocity fields in the Fourier space. Now, for completeness of our considerations, in this section we will focus on the configuration space. The clustering statistics of quantities defined in the configuration space provides a complementary
picture of the field properties. The variance and the two-point correlation functions of a cosmic field are related to its Fourier power spectrum by
\bq
\label{eqn:xi2-pk}
\xi(r)=\int{{\rm d} k\over2\pi^2} k^2 P(k){\sin(kr)\over(kr)}\,,\\
\label{eqn:sigma-pk}
\sigma^2(r) = \int{{\rm d} k\over2\pi^2} k^2 P(k)W_{\rm TH}^{2}(kr)\,.
\eq
Here $W_{\rm TH}$ is the Fourier top-hat window and $r$ is the comoving separation (or smoothing) scale in $\hmpc$. We have computed both variance and 
two-point correlation function for the density and velocity divergence fields for all our $L_{256}$ runs. For a set of smoothing scales satisfying
$1\leq r/(\hmpc)\leq 0.1L_{box}$ the denoted differences between quasi-static and time derivatives runs were even smaller then any of the differences
we have observed for the density and velocity power spectra  shown in figures from \ref{F4lowRes} to \ref{velPS}. Thus we can conclude that both 
frequency and configuration space two-point statistics used so far in this study are fostering consistent picture. This reassures us that any differences
in the properties of the density and velocity fields between quasi-static and time derivatives runs must be very small.

\cite{2010PhRvD..82j3536H, 2013MNRAS.435.2806H} have indicated that the high-order moments are much more sensitive probes of even minute
changes in the cosmic density field. They have shown in particular, that the clustering amplitudes are well posed to emphasise even very small
differences in the clustering pattern when applied for modified gravity models. Following method of~\cite{2013MNRAS.435.2806H}, we have computed the reduced skewness
$S^{\delta;\theta}_3\equiv \langle\delta^3;\theta^3\rangle\sigma_{\delta;\theta}^{-4}$ and the reduced kurtosis 
$S^{\delta;\theta}_4\equiv \langle\delta^4;\theta^4\rangle\sigma_{\delta;\theta}^{-6}- 3\sigma_{\delta;\theta}^{-2}$ for our ensemble of $L_{256}$ simulations.
For all the relevant smoothing scales we have not found any significant differences between quasi-static and time derivatives realisations in any of our runs.

The results described in Section~\ref{sec:PS} augmented by our findings concerning the configuration space clustering statistics clearly demonstrate, that
in the statistical sense the cosmic density and velocity fields produced in quasi-static and time derivative runs are equivalent down to resolved scales.

\begin{figure}
\centering
\includegraphics[trim = 0mm 0mm 0mm 0mm, clip,width=\columnwidth, height = 0.4\textheight]{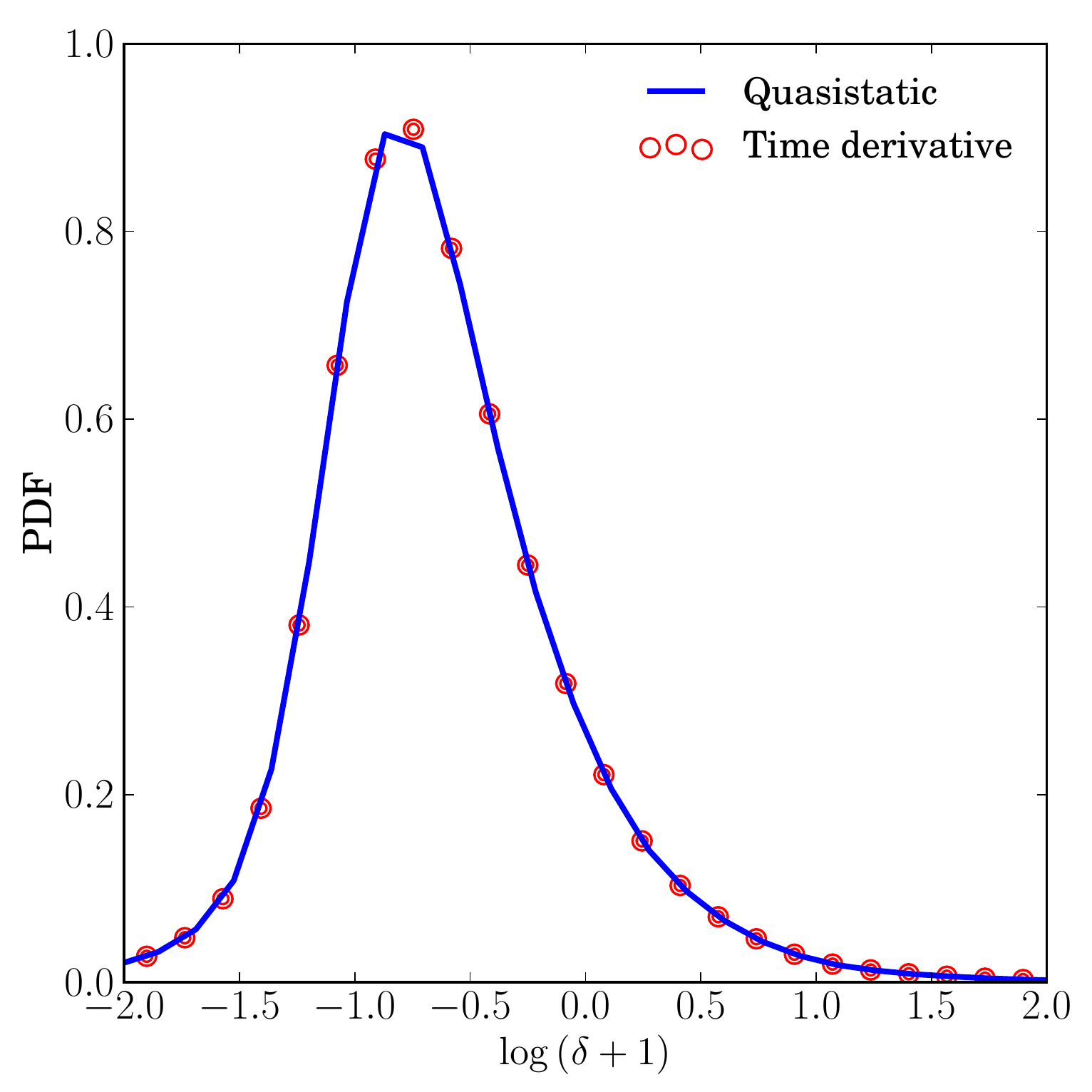}
\caption{Probability density functions for the density field $\delta + 1$ in F6 at $z = 0.5$ computed within a spherical top-hat window, smoothed at $r = 0.25~h^{-1}\,\mathrm{Mpc}$. The distribution for the quasi-static simulation is shown in the blue line, whereas that of the non-static simulation is displayed in the red circles.}
\label{densityPDF}
\end{figure}

Finally, to summarise this section we show in Fig.~\ref{densityPDF} the probability distribution functions (PDFs) of the density field
computed at $z=0.5$ for our high-resolution $L_{64}$ runs. Here we compare only the PDFs of the F6 brand modelled in our two approaches, with the smoothing scale, $r = 0.25~h^{-1}\,\mathrm{Mpc}$ (equivalent to the size of one grid cell in $L_{64}$). Comparing the PDFs
of the two realisations serves as our final test. So far we have focused on statistical quantities, in which any signal coming from relatively small spatial
regions would be strongly suppressed. One could imagine that there might exist some special regions in the density field, where the time derivatives 
of the scalaron could take bigger values and hence make a bigger impact the dynamics of the cosmic fields. The very centres of cosmic voids can serve as one 
example of such a place. The extremely low density in those locations could in principle allow for much stronger non-linear behaviour of the scalar field.
However, analysis of the data plotted in the Fig.~\ref{densityPDF} evidently convinces us that both high and low $\delta$ tails of the PDF agree remarkably
well in the compared simulations. All extreme objects, like very deep voids or very massive clusters, populate the aforementioned tails of the density PDF.
The fact that the both curves agrees also in these regions guarantee that the scalaron and the matter fields exhibit the same dynamical evolution
in both quasi-static and time derivative simulations.

\section{Numerical Considerations}
\label{NumCons}

In this section, we discuss some of the code-specific numerical issues that one needs to account for in the \tcr{interpretation of our results above}. 

\subsection{Convergence of Solutions}
\label{Conv}

In {\sc ecosmog}, between successive \tcr{relaxation sweeps,} 
one can define a residual $d^h$,  as the \tcr{difference between the numerical values of the} two sides of the equations being solved. Convergence (or alternatively, the signal to ``stop'' further 
\tcr{relaxation iterations}) is achieved when the residual gets smaller than \tcr{some predefined threshold, the so-called {\it convergence criterion}. In practice, however, the accuracy one would ever achieve when numerically solving our partial differential equation is fundamentally limited by a numerical error, the so-called truncation error $\tau^h$, imposed by the discretisation of the continuous differential equation. The latter implies that there is no point to further reduce the residual by doing more relaxation iterations, once it has become smaller than the truncation error \cite{2002nrc..book.....P}}:
\bq \label{criterion}
\left| d^h \right| \leq \alpha \left| \tau^h \right|\;,
\eq
where $\alpha$ is some constant ($\sim 1/3$). 

Throughout \tcr{this} work, convergence is \tcr{deemed to have been} achieved when the residual $|d^h| \leq 10^{-8}$, which is a significantly stronger criterion than that in Eq.~\ref{criterion}, and \tcr{further reducing $|d^h|$} does not change the results by much. If, however, one uses $\left| d^h \right| \leq \alpha \left| \tau^h \right|$, then the results will be changed, and the change itself is larger than the offsets \tcr{caused by including} the time derivatives. \tcr{Obviously}, this is a change that we have no control over. The quasi-static approximation therefore introduces an error well below that caused by the discretisation \tcr{of the differential equation itself} \footnote{\tcr{It is often argued that one should make $|d^h|$ as small as practically possible, instead of stopping at $|d^h|\sim\tau^h/3$, to prevent the numerical errors in solving the differential equation at individual steps from accumulating over the many time steps of a simulation. While this is true to certain extent, it is 
not clear that the discretisation error itself will not accumulate in this case (recall that, if $d^h$ could be brought to zero, then the remaining error is completely from the discretisation). Again, the way to get away from this problem is to reduce the discretisation error by increasing the (spatial) resolution, and then check for convergence.}}.   

\subsection{Box Size and Resolution}
\label{Box}

As we have seen in the previous section, the results that we get for the F6 simulations depend on the resolution. This is quite an odd result, and on first instance, slightly contrary to what one might expect when considering the CFL condition in Eq.~\ref{CFLcode}, which is what {\sc ecosmog} uses to determine the size of the integration time step. Reducing the box size (as we have done here) will reduce $\mathrm{d} x$ by the same factor, but the integration time step is also affected in the same way to ensure that particles do not move more than a cell size during one time step. As such, one would expect that adjusting the resolution by means of a increase or decrease in the size of the simulation box should not affect how different the time derivative case is from the quasi-static limit. Why then do the higher-resolution $L_{128}$ (Fig.~\ref{F6highRes}) and $L_{64}$ runs reduce the discrepancy of this offset seen in the $L_{256}$ run (Fig.~\ref{F6lowRes})?

As a result of \tcr{decreasing the} time step in the higher resolution simulations, the particles do not \tcr{travel as far as they do in low-resolution simulations} in a given time step, and so between two consecutive steps, the $f_R$ field configuration in real space does not change \tcr{as much} as in a full time step run, which makes its time derivatives smaller. In terms of the equations of motion (say in Eq.~\ref{discfR}), this amounts to saying that the value $u_{i,j,k}^{(n)} - u_{i,j,k}^{(n-1)}$ does not change as much when the time step is reduced. We have tested this in the $L_{256/2}$ run, by re-running the $L_{256}$ simulation in F6, this time artificially halving the time step that {\sc ecosmog} would naturally use (keeping the force resolution the same), and find that the offset between the quasi-static and time derivative runs is indeed reduced \tcr{as in the $L_{128}$ simulations}. 

We thus \tcr{conclude} that with increased resolution, the reduced time steps \tcr{make} the quasi-static and non-static F6 simulations converge better, and 
in the convergence limit the time derivatives do not have a big impact \tcr{on any of our $f(R)$ gravity simulations}.

\subsection{Inaccuracies Due to Averaging Over Oscillations}
\label{avgerrors}

One of the major caveats behind our analysis is the manner in which we 
\tcr{include} the time evolution of $f_R$ in our simulations. Since $m_{\mathrm{eff}} \gg H$, the scalaron field is expected to oscillate very fast about its minimum as it evolves. As mentioned in Section~\ref{Disc}, the time derivative is calculated by averaging $f_R$ over the many oscillations. We can make a crude estimate of the error \tcr{caused by} this procedure by following the methodology of \cite{2012PhRvD..86d4015B}. 

The background evolution of the field scalaron is given by the equation:
\bq \label{backEv}
\ddot{f}_R + 3 H \dot{f}_R + \frac{\mathrm{d} V_{\mathrm{eff}}}{\mathrm{d} f_R} = 0\;,
\eq
where $\mathrm{d} V_{\mathrm{eff}} / \mathrm{d} f_R = \tcr{-}1/3 \left(R - f_R R + 2 f(R) + 8\pi G \rho_m\right)$ \tcr{as defined in Eq.~\ref{effPot}}. Now, \tcr{let us} consider \tcr{small} perturbations of the scalaron about its minimum as $\delta f_R \equiv f_R - f_{R, \mathrm{min}}$ \tcr{(note that across this subsection $\delta f_R$ is {\it not} the spatial perturbation)}, \tcr{and derive the following evolution equation for $\delta f_R$:}
\bq
\delta \ddot{f}_R + 3 H \delta \dot{f}_R + \tcr{m_{\rm eff}^2} \delta f_R &=& F\left(t\right) \nonumber \\  
&\equiv& -\frac{1}{a^3} \frac{\mathrm{d}}{\mathrm{d} t} \left[ a^3 \frac{\mathrm{d} f_{R,\mathrm{min}}}{\mathrm{d} t}\right].
\eq

The minimum equation for $f(R)$ gravity is given by:
\bq
\left. \frac{\mathrm{d} V_{\mathrm{eff}}}{\mathrm{d} f_R} \right|_{f_{R,\mathrm{min}}} = 0\;,
\eq
which \tcr{has been used to derive the above equation and which also} implies that \tcr{(by taking the time derivative of the relation $R\approx-8\pi G\rho_m$)}:
\bq
\frac{{\rm d} {f}_{R,{\rm min}}}{{\rm d}t}\ \tcr{\approx}\ -\frac{\tcr{8}\pi G \rho_m}{m_{\mathrm{eff}}^2}H\;.
\eq

The time-dependent force term $F(t)$ then becomes:
\bq
F\left(t\right)\ \tcr{\approx}\ \frac{\tcr{8}\pi G \rho_{m0}}{a^3} \frac{\mathrm{d}}{\mathrm{d} t} \left[ \frac{H}{m_{\mathrm{eff}}^2} \right]\;.
\eq

In addition to \tcr{being driven by the (slow) time evolution of} the minimum \tcr{$f_{R,{\rm min}}$}, the scalaron field 
\tcr{also experiences a number of ``kicks''} when relativistic species 
\tcr{become non-relativistic and thus starts to contribute to $T^\mu_{\ \mu}\approx\rho_m$}. 
\tcr{Because the transition from relativistic to non-relativistic happens on a relatively short time scale compared to the Hubble time, we can model this effect as ``instantaneous kicks'' \cite{2004PhRvD..70l3518B}\footnote{\tcr{The kick is by the sudden increase in the non-relativistic $\rho_m$, as can be seen from $\mathrm{d} V_{\mathrm{eff}} / \mathrm{d} f_R = -\left[R - f_R R + 2 f(R) + 8\pi G \rho_m\right]/3$ -- because of the quick change in $\rho_m$, $f_{R,{\rm min}}$ is changed while the true $f_R$ needs time to respond to this.}}}:
\bq
F\ \tcr{\approx}\ \frac{\tcr{8} \pi G \rho_{m0}}{a^3} \frac{\mathrm{d}}{\mathrm{d} t} \left[ \frac{H}{m_{\mathrm{eff}}^2} \right] - \tcr{\beta}\sum_j \kappa_j H_j \delta \left( t - t_j \right),
\eq
where $t_j$ is the time \tcr{at which the transition from relativistic to non-relativistic happens} and $\kappa_j \approx g / g_\star \left(m_j\right) \leq 1$, with \tcr{$g$ the number of degrees of freedom of the species that is becoming non-relativistic and} $g_\star \left(m_j\right)$ the number of relativistic species at \tcr{time $t_j$, when the temperature $T$ is equal to the mass $m_j$. $\beta\sim\mathcal{O}(1)$ is a constant and $H_j$ is the Hubble expansion rate at $t_j$}. 

In what follows, we limit ourselves to the time of the electron decoupling, $t_e$\tcr{, as an example of the analysis}:
\bq
\delta \ddot{f}_R + 3 H \delta \dot{f}_R + \tcr{m^2_{\rm eff}} \delta f_R &\tcr{~\approx~}& \frac{\tcr{8}\pi G \rho_{m0}}{a^3} \frac{\mathrm{d}}{\mathrm{d} t} \left[ \frac{H}{m_{\mathrm{eff}}^2} \right] \nonumber \\ 
&& \tcr{- \beta_e} \kappa_e H_e \delta \left( t - t_e \right)\;.
\eq
\tcr{Defining a new field $\psi$ which satisfies} $\delta f_R = a^{-3/2} \psi$, \tcr{this equation can be rewritten as}:
\bq \label{psiEq}
\ddot{\psi} + \left[ \tcr{m^2_{\rm eff}} + \frac{9}{4} w H^2 \right] \psi &\tcr{~\approx~}& \tcr{8}\pi G \rho_{m0} a^{-3/2} \frac{\mathrm{d}}{\mathrm{d}t} \left[\frac{H}{m_{\mathrm{eff}}^2}\right] \nonumber \\ 
&& \tcr{-\beta_e}  \kappa_e H_e a_e^{3/2} \delta \left(t - t_e\right),
\eq
where \tcr{$w=P/\rho$ is the effective} equation of state, \tcr{with $\rho, P$ including contributions from all matter species}. Since $m^2 \gg H^2$, we can solve Eq.~\ref{psiEq} using the Wentzel-Kramers-Brillouin (WKB) approximation, and finally get:
\bq
&&\delta f_R\ \tcr{\approx}\ \frac{\tcr{8}\pi G \rho_{m0}}{m_{\mathrm{eff}}^2 a^3} \frac{\mathrm{d}}{\mathrm{d} t} \left[ \frac{H}{m_{\mathrm{eff}}^2} \right] \\ && -\Theta \left(t -t_e\right) \tcr{\beta_e} \kappa_e H_e \frac{a_e^{3/2}}{a^{3/2}} \frac{1}{\sqrt{m_e m_{\mathrm{eff}}}} \sin \int^t_{t_e} m\left(t'\right) \mathrm{d} t'\;. \nonumber
\eq
where $\Theta\left( t-t_e \right)$ is the Heaviside function, \tcr{$m_e \equiv m_{\rm eff}(t=t_e)$, $H_e=H(t=t_e)$ and similarly for $\beta_e$ and $\kappa_e$.} 

By rewriting
\bq
\frac{\mathrm{d}}{\mathrm{d} t} \left[ \frac{H}{m_{\mathrm{eff}}^2} \right] = g (t) \frac{H\tcr{^2}}{m_{\mathrm{eff}}^2}\;,
\eq
\tcr{and using}
\bq
\tcr{8\pi G\rho_{m0}\ =\ 3H_0^2\Omega_m,} 
\eq
we can finally average over the rapid oscillations to get:
\bq
\langle \delta f_R^2 \rangle\tcr{(t)}\ &\tcr{\approx}&\ \tcr{\frac{9\Omega_m^2g^2(t)}{a^6}\frac{H_0^4}{m_{\rm eff,0}^4}\frac{m_{\rm eff,0}^4}{m^4_{\rm eff}(t)}\frac{H^4}{m^4_{\rm eff}(t)}}\nonumber\\
&& \tcr{+ \frac{\beta^2_e\kappa_e^2}{2}\frac{a^3_e}{a^3}\frac{H_e^2}{m^2_e}\frac{m_e}{m_{\rm eff}(t)}},
\eq
\tcr{where, again, a subscript $_0$ denotes the value at present day}.

\tcr{At late times, e.g., $a\sim1$, the first term in the above expression is of order $(H_0/m_{\rm eff,0})^8$ and is extremely small (compared to $\left|f_{R0}\right|$) because $H_0/m_{\rm eff,0}$ is typically less than $10^{-3}$ for the models studied here. This term appears because of the shift of $f_{R,{\rm min}}$, which itself is due to the evolution of the background matter density in the Universe. It has nothing to do with the oscillations that we are interested in here.}

\tcr{The second term characterises the amplitude of the oscillations of $\delta f_R$. Up until the onset of the acceleration phase, we have $|\bar{f}(R)|\ll8\pi G\bar{\rho}_m$ and therefore $\bar{R}\approx-8\pi G\bar{\rho}_m$, where $\bar{\rho}_m$ has no contribution from radiation even in the radiation-dominated era. This relation $m^2_{\rm eff}(t)\approx-(1/3){\rm d}R/{\rm d}f_R$ gives}
\bq
\tcr{m^2_{\rm eff}(t)}\ &\tcr{\approx}&\ \tcr{\frac{H_0^2\Omega_m}{3n(n+1)\xi}\left[\frac{-R}{M^2}\right]^{n+2}}\nonumber\\
&\tcr{\approx}& \tcr{~3^{n+1}\frac{H_0^2\Omega_m}{n(n+1)\xi}a^{-3(n+2)}.}
\eq 
\tcr{By noting that $n=1$, $3^{n+1}\Omega_m\sim\beta_e^2\kappa_e^2\sim\mathcal{O}(1)$, we can combine the above two equations to estimate the amplitude of the oscillation as
\bq
\langle\delta f_R^2\rangle^{1/2}(t) \sim \xi a_e^{7/2}\Omega_{r}a^{3/2}
\eq
where $\xi\approx34^2|\bar{f}_{R0}|$, $\Omega_r\sim10^{-4}$ is the present-day fractional energy density of radiation and $a_e\sim10^{-9}$ is the scale factor at $t_e$. The late-time dominance of dark energy slightly alters the relation $\bar{R}\approx-8\pi G\bar{\rho}_m$, but nevertheless the above result still serves as a good order-of-magnitude estimate.}

\tcr{We are more interested in the quantity
\bq
\frac{\langle\delta f_R^2\rangle^{1/2}(t)}{|\bar{f}_{R,{\rm min}}(t)|} \sim 9a_e^{7/2}\Omega_{r}a^{-9/2},
\eq
which is independent of $|\bar{f}_{R,0}|$ and decays over time. A quick calculation shows that for our simulations ($z<49$) the amplitude of the oscillation is always smaller than $10^{-27}$ times $\bar{f}_{R,{\rm min}}$, with a value of $10^{-35}$ today\footnote{\tcr{Note that we can use $a_e$ in the above expressions and estimates, because electrons are the last species of standard-model particles that become non-relativistic.}}.}

Evidently, with such tiny amplitudes, the oscillations are unlikely to have any impact on our result, and the averaging over many oscillations should work accurately. Note also that the smallness of $\langle\delta f_R^2\rangle^{1/2}(t)$ implies that it is probably unrealistic to follow the oscillations using explicit time integration in a numerical simulation poised for the study of cosmic structure formation, such as ours here.

Of course, the analysis in this subsection has been greatly simplified. In reality, the situation could be much more complicated. For example, the scalaron field $f_R$ at a given position of space may not be oscillating around the minimum of its effective potential as determined by matter density at that position, but instead far from that minimum due to interactions with the density field in the environment; the oscillations could have a position (or local-density) dependent mass $m_{\rm eff}(t,\vec{x})$; and there can even be `micro kicks' caused by rapid changes 
of local matter density due to particles moving to or away from the position, etc.. Such `micro kicks' may not be well approximated as instantaneous kicks because particle velocity $v\ll c$, and they have already been accounted for in our time integration scheme.


\subsection{Initial Conditions}
\label{ICs}

We see from Fig.~\ref{F4lowRes}-\ref{F6highRes} that the different initial conditions 
can lead to significant \tcr{variations} in the results. This can be seen in the form of error bars on the data points in the figures, which represent the scatter within each $k$-bin over the five realisations -- the relative enhancement of the power spectra $\Delta P/P$ can be lower or higher than the mean 
of the bin. Our results demonstrate that the variations across different realisations dominate the differences induced by including time derivatives.

\subsection{The Effect of Baryons}
\label{Baryons}

In this paper, we have ignored the effect of baryons in our simulations. While this is not expected to make much of a difference on large scales, the baryonic effects are more pronounced on nonlinear scales, making it more difficult to correctly measure the power spectrum $P(k)$ in this regime. \cite{2014MNRAS.440.2997V} found that there can be a discrepancy of more than 10\% in the two-point correlation function on sub-Mpc scales between dark matter only simulations, and those with baryonic effects included. The difference between the inclusion and non-inclusion of time derivatives in our $f(R)$ gravity simulations is typically sub-percent, so we expect that any errors from the non-inclusion of baryons significantly dominate those caused by the quasi-static approximation.

\section{Discussion and Summary}
\label{Summary}

In this paper, we have studied the effect of including time derivatives in the scalar field equation of motion in \tcr{numerical simulations of structure formation for }$f(R)$ gravity, which is a departure from the 
\tcr{quasi-static approximation usually used in such simulations.} 
To this end, we have \tcr{generalised} both the $f(R)$ equation itself (Section~\ref{MEOM}) and the Poisson equation (Section~\ref{modPE}), which are the equations that govern the formation of cosmic structures in this model. We find that, 
\tcr{in both cases, the inclusion of time derivatives results in additional terms entering the equations compared to} the quasi-static case, as seen in Eq.~\ref{FullEqs}. 
To solve these equations, we make use of {\sc ecosmog}, using $256^3$ particles in different boxes (of size $256\,h^{-1}$ Mpc, $128\,h^{-1}$ Mpc and $64\,h^{-1}$ Mpc), to test for the effects of resolution. In the low-resolution case, we evolve three different Hu-Sawicki $f(R)$ models: F4, F5 and F6, corresponding to different values of the scalaron field $\left|\bar{f}_{R0}\right|$ (Section~\ref{Results}).

By looking at the enhancement of the matter and velocity divergence power \tcr{spectra} relative to $\Lambda$CDM, we find that, in the \tcr{cases} of F4 and F5 (Section~\ref{lowRes}), the low-resolution $L_{256}$ box \tcr{simulations confirm} that including time derivatives introduces only \tcr{an insignificant} difference from the quasi-static approximation, whereas this difference is larger in the case of F6. To see if changing box size has any effect on this discrepancy, we perform the F6 simulations in the $L_{128}$ runs (Section~\ref{highRes}), and find that this large offset \tcr{becomes smaller}. 
\tcr{To verify whether this is actually a consequence of increasing the resolution}, we also run \tcr{two additional tests. The first is} a variation of the original $L_{256}$ simulation but \tcr{with its time steps artificially halved} (which we dub the $L_{256/2}$ simulation). \tcr{This simulation has the same mass and force resolution as the low-resolution $L_{256}$ runs, but it shows the same decrease of the non-static effect as in $L_{128}$. The second is an ever higher resolution simulation with $256^3$ particles in a box of size $64\,h^{-1}$Mpc, which we call $L_{64}$. This simulation has even smaller time steps and shows even better agreement between the quasi-static and non-static runs. Finally, we test the statistics of the configuration space for both the static and non-static cases, and again find no discernible differences.}

The implications of the additional tests are twofold: 
\begin{enumerate}[i]
\item They confirm that with increasing temporal resolution, our implicit scheme for time integration does converge, and this is a nontrivial check that our new code and algorithm works consistently;
\item The converged result is that, even for F6, the inclusion of time derivative is neither crucial nor necessary, and that the quasi-static approximation works reasonably well for all $f(R)$ models studied here..
\end{enumerate}

We have also discussed  numerical issues associated with our algorithm. In particular, our time-integration scheme assumes implicitly that the code actually evolves quantities which are averaged over many scalaron field oscillations. Our qualitative analysis shows that the amplitudes of such oscillations, although grow in time, are much smaller than the average value (i.e., the oscillation centre) at all epochs of interest to us, and as a result the implicit time-average should have no impact on our result in practice. We have also discussed other intrinsic sources of scatter, such as the different initial conditions (cosmic variance) and the convergence criterion for our relaxation method, and concluded that they are all significantly larger than the possible error caused by the quasi-static approximation.


To summarise: we find that the effects of the scalar field time derivatives are so small that can be safely neglected for the most practical applications in cosmology.

The three models we consider -- F4, F5 and F6 -- span a wide range in the strength of the screening mechanism, from very weak to very strong, 
\tcr{but in all these cases} the quasi-static approximation holds yielding reliable results. \tcr{In particular, F4 corresponds to a model where the chameleon screening is so weak that it is closer to unscreened theories such as coupled quintessence \cite{2011PhRvD..83b4007L,2011MNRAS.413..262L}, and the conclusion can be generalised to those classes of theories.}

\tcr{On the other hand, we must be cautious when trying to generalise the conclusion here to other modified gravity theories. An important example is the Galileon gravity model \cite{2009PhRvD..79f4036N,2009PhRvD..79h4003D}, which has the Dvali-Gabdadze-Porrati (DGP) model \cite{2000PhLB..485..208D} as a subclass. \cite{2013JCAP...10..027B,2013JCAP...11..012L} found that neglecting the time derivatives results in the equations having no real solutions in low-density regions, which does not occur in the case of $f(R)$ gravity. As a result, for those theories, the time derivatives are likely to have a non-negligible impact on the cosmic fields. It would be interesting to apply our method of including non-static effects to Galileon simulations and quantify this impact, and this will be left for future work.}

\begin{acknowledgments}

We would like to thank Alexandre Barreira for his help in generating initial conditions, and for other useful comments. \tcr{We would also like to thank the participants of the MPA modified gravity workshop organised by Fabian Schmidt, for useful discussions. SB is supported by STFC through grant [ST/K501979/1,ST/L00075X/1]. BL is supported by the Royal Astronomical Society and Durham University.} WAH appreciates support from ERC Advanced Investigator grant COSMIWAY 
[grant number GA 267291] and the Polish National Science Center [grant number DEC-2011/01/D/ST9/01960].
This work used the DiRAC Data Centric system at Durham University, operated by the Institute for Computational Cosmology on behalf of the STFC DiRAC HPC Facility (\url{www.dirac.ac.uk}). This equipment was funded by BIS National E-infrastructure capital grant ST/K00042X/1, STFC capital grant ST/H008519/1, and STFC DiRAC Operations grant ST/K003267/1 and Durham University. DiRAC is part of the National E-Infrastructure. 
\end{acknowledgments}


\bibliography{bibliography.bib}

\end{document}